\documentclass{article}

\usepackage[affil-it]{authblk}
\usepackage{amsmath}
\usepackage{amstext}
\usepackage{amssymb}
\usepackage{mathrsfs}
\usepackage{graphicx}
\usepackage{envmath}
\usepackage{subfig}

\newcommand{\D}{\ensuremath{\mathcal{D}}}
\newcommand{\M}{\ensuremath{\mathcal{M}}}
\newcommand{\Ms}{\ensuremath{\widetilde{\mathcal{M}}}}
\newcommand{\Ss}{\ensuremath{\mathcal{S}}}
\newcommand{\Sp}{\ensuremath{\mathcal{S}_p}}

\newcommand{\MTT}{\ensuremath{\mathcal{H}}} 
\newcommand{\Lie}{\ensuremath{\mathscr{L}}}
\newcommand{\rhot}{\ensuremath{\widetilde{\rho}}}

\begin{document}

\title{Quasi-local black hole horizons}
\author{Badri Krishnan}
\affil{Max Planck Institute for Gravitational Physics (Albert Einstein Institute),
Callinstr. 38, 30167 Hannover, Germany}
\maketitle


\abstract{ This article introduces the subject of quasi-local horizons
  at a level suitable for physics graduate students who have taken a
  first course on general relativity.  It reviews properties of
  trapped surfaces and trapped regions in some simple examples,
  general properties of trapped surfaces including their stability
  properties, the definitions and some applications of dynamical-,
  trapping-, and isolated-horizons.  }


\section{Introduction}
\label{sec:intro}

The first conception of a black hole was due to Michell and Laplace in
the $18^{\rm th}$ century.  They viewed it as a star whose
gravitational field is so strong that the Newtonian escape velocity
$\sqrt{2GM/R}$ (with $M$ and $R$ being the mass and radius of the star
respectively) is larger than the speed of light. The condition on the
escape velocity leads to the inequality $R\leq 2GM/c^2$ which,
remarkably, also holds in general relativity.  While such a star would
have the property that not even light can escape from it, this is
however a non-relativistic concept. The speed of light is not
privileged in pre-relativistic physics, and a moving observer would
not necessarily see it as a dark object.  A more complete account of
this history is given by Hawking and Ellis \cite{HawkingEllis:1973}
including reprints of the original articles.

The history of black holes proper dates back to just after the
discovery of general relativity.  The first non-trivial exact solution
to the Einstein equations discovered by Schwarzschild in 1916 and
named after him was, in fact, a black hole.  It was however more than
four decades before its properties were fully appreciated.  The
Kruskal-Szekeres extension of the Schwarzschild solution was
discovered only in 1960.  This was shortly followed by the discovery
of the Kerr solution in 1963 representing spinning black holes. Its
global properties were explained by Carter in 1966.  The Kerr-Newman
solution representing charged, spinning black holes was discovered in
1965.  It was in 1964 that the phrase ``black hole'' was first coined
by John Wheeler.  During the same time, there were seminal
developments in understanding the general properties of black holes
beyond specific examples.  This includes the study of the global
properties of black hole spacetimes, the definition of event horizons,
and crucially for the developments to be discussed here, the
singularity theorems of Penrose and Hawking and the introduction of
trapped surfaces by Penrose.  This was soon followed by the
understanding of black hole thermodynamics by Bekenstein, Bardeen,
Carter and Hawking in 1973, and the discovery of Hawking radiation in
1974.  The cosmic censorship hypothesis was formulated by Penrose in
1979.  The question of whether this is valid, i.e. if every
singularity that results from the future evolution of generic, regular
initial conditions is hidden behind an event horizon, is not settled
and is one of the key unsolved questions in classical general
relativity.  The black hole uniqueness theorems which showed that the
Kerr-Newman solutions are the unique globally stationary black hole
solutions in Einstein-Maxwell theory in four dimensions was
established in the 1980s following the work of Israel, Carter and
Robinson.

More recently, black holes have been the subject of intense study in
quantum gravity where the calculation of black hole entropy has been
seen as a key milestone for string theory and loop quantum gravity.
There have also been important developments on the classical side
where the long standing problem of calculating the gravitational wave
signal from the the merger of two black holes was finally solved
numerically in 2005 by Pretorius.  In an astrophysical context, black
holes are believed to be engines for some of the most violent events
in out universe, such as active galactic nuclei.  Astronomers have
succeeded in locating a large number of black hole candidates with
masses ranging from a few to billions of solar masses, and the direct
detection of gravitational waves from binary black hole systems is
expected later this decade.

Most of these seminal developments have relied on event horizons to
characterize the boundary of the black hole region (the singularity
theorems are a notable exception).  This is completely reasonable when
we are dealing with stationary situations, but can lead us astray in
dynamical situations.  As we shall elaborate later, event horizons are
global notions and it is in principle not possible for a mortal
observer to locate them.  One possible alternative is to use the
notion of trapped surfaces introduced by Penrose.  While not entirely
local since they are closed spacelike surfaces, these provide a
quasi-local alternative which an observer could in principle locate in
order to detect the presence of a black hole.  Trapped surfaces lead
logically to various kinds of quasi-local horizons including isolated,
dynamical and trapping horizons. The goal of this chapter is to
motivate and explain the quasi-local approach to studying black hole
horizons, and to review some recent results.  Somewhat surprisingly,
we shall see there is still a major gap in our understanding of
classical black holes in dynamical spacetimes.  If we accept black
holes as bonafide astrophysical objects, we still do not have a
satisfactory notion of what the surface of a black hole is. Event
horizons are not satisfactory because of their global properties, but
there is as yet no established quasi-local alternative.

The style of this chapter will typically be to start informally with
simple examples and to use them as guidance for developing general
concepts and definitions.  In Sec.~\ref{sec:examples} we shall start
with the simplest black hole, i.e. the spherically symmetric
Schwarzschild spacetime, and understand the properties of its black
hole region.  This naturally motivates the fundamental notions of
event horizons and trapped surfaces, and to the boundary of the
trapped region.  As we shall see, the different reasonable definitions
of the black hole horizon agree in Schwarzschild.  This will not be
the case in more general situations.  Perhaps the simplest example is
the imploding Vaidya spacetime which shall be our second example in
Sec.~\ref{subsec:vaidya}.  We shall see that at least in this simple
spherically symmetric example, the location of the trapped surfaces
can be determined.  These two examples are then followed by general
definitions of event horizons, and trapped surfaces in
Sec.~\ref{sec:definitions} which formalizes many notions introduced in
Sec.~\ref{sec:examples}.  Some properties of trapped surfaces under
deformations and time evolution are then discussed in
Sec.~\ref{subsec:stability}, and this leads naturally to the notions
of marginally trapped tubes, and trapping and dynamical horizons.  We
then restrict our attention to isolated horizons which describes the
equilibrium case, when no matter or radiation is falling into the
black hole (but the rest of spacetime is allowed to be dynamical).
This is now well understood and we review the general formalism in
Sec.~\ref{sec:ih} with the Kerr black hole as the prototypical
example.  In particular, we discuss two applications: black hole
thermodynamics and the spacetime in the neighborhood of an isolated
black hole.  Sec.~\ref{sec:dh} reviews some results and applications
for dynamical horizons and finally Sec.~\ref{sec:outlook} provides a
summary and some open issues.

The discussion of this chapter will be mostly self-contained, though
digressions into the relevant mathematical concepts will be
necessarily brief.  Useful references for black holes and general
relativity are \cite{Chandrasekhar:1985kt,Wald:1984}, and for more
mathematically inclined readers \cite{Lee:1997} is recommended as a
useful introduction to the relevant concepts in differential geometry.
The discussion is meant to be accessible to physics graduate students
who have taken a first course in general relativity (covering, say,
the first part of the textbook by Wald \cite{Wald:1984}).  In the same
spirit, the list of references is not meant to be exhaustive in any
sense, and is mostly biased towards reviews and pedagogical
material. The selection of topics is not meant to be exhaustive; this
contribution is \emph{not} to be viewed as a broad review article.  It
is rather a combination of pedagogically useful examples, and a brief
description of some recent results.  This material will hopefully whet
the reader's appetite and motivate him/her to delve further into the
subject.

Some words on notation are in order.  A \emph{spacetime} is a smooth
four-dimensional manifold $\mathcal{M}$ with a Lorentzian metric
$g_{ab}$ with signature $(-+++)$.  We shall use a combination of
index-free notation and Penrose's abstract index notation for tensors
\cite{Penrose:1985jw}; lower-case Latin letters $a,b,c,\ldots$ will
denote spacetime indices.  Symmetrization of indices will be denoted
by round brackets, e.g. $X_{(ab)} := (X_{ab} + X_{ba})/2$, and
anti-symmetrization by square brackets: $X_{[ab]} := (X_{ab} -
X_{ba})/2$.  The derivative-operator compatible with $g_{ab}$ will be
denoted $\nabla_a$, and the Riemnann tensor $R_{abcd}$ will be defined
by $2\nabla_{[a}\nabla_{b]}X_c = {R_{abc}}^d X_d$ for an arbitrary
1-form $X_a$.  The Ricci tensor and scalar are respectively
$R_{ab}={R_{acb}}^c$ and $R=g^{ab}R_{ab}$. The coordinate derivative
operator will be denoted $\partial$.  The exterior derivative will be
either denoted by indices, such as $\nabla_{[a}X_{b]}$, or in index
free notation as $dX$.  The Lie derivative of an arbitrary tensor
field $T$ along a vector field $X$ will be denoted $\mathscr{L}_XT$.
Where no confusion is likely to arise, we shall often not explicitly
include the indices in geometric quantities.  Unless otherwise
mentioned, we shall work in geometrical units with $G=1$ and $c=1$.
We shall often deal with sub-manifolds of $\mathcal{M}$; a
sub-manifold of unit co-dimension will be called a hyper-surface while
lower dimensional manifolds (typically these will be 2-spheres
topologically) will be called surfaces. All sub-manifolds shall be
assumed to be sufficiently smooth.  Unless mentioned otherwise, we
shall be working with standard general relativity in four spacetime
dimensions.

\section{Simple examples}
\label{sec:examples} 

\subsection{The trapped region in Schwarzschild spacetime}
\label{subsec:schwarzschild}

We shall start by studying the gravitational field in the vicinity of
a time-independent massive spherically-symmetric body.  In this
section we recall some basic properties of the Schwarzschild solution.

The Schwarzschild metric is a static, spherically symmetric solution
of the vacuum Einstein equations $R_{ab} = 0$.  It is usually presented
as
\begin{equation}
  \label{eq:sch-standard}
  ds^2 = -\left(1-\frac{2M}{r}\right)dt^2 + \left(1-\frac{2M}{r}\right)^{-1}dr^2 
  + r^2(d\theta^2 + \sin^2\theta\,d\phi^2)\,. 
\end{equation}
Here $r$ is a radial coordinate such that the area of spheres at fixed
$r$ and $t$ is $4\pi r^2$; these spheres can be obtained invariantly
by applying rotational isometries to a given initial point in the
manifold.  Each of these spheres are isometric to the standard round
spheres in Euclidean space, and $(\theta,\phi)$ are the usual polar
coordinates.  The time coordinate is $t$, and the metric is explicitly
time independent in these coordinates.  The parameter $M$ is the mass
and in non-geometrical units, we would have the replacement
$M\rightarrow GM/c^2$.  This metric turns out to be an excellent
approximation to, say, the gravitational field in our solar system
with the sun treated as a point mass and with $\phi(r) = GM/c^2r$
being its Newtonian gravitational potential. The quantity $R_s :=
2GM/c^2$ is known as the Schwarzschild radius, and for the sun $R_s
\approx 3\,$km (which agrees with the Michell-Laplace idea mentioned
at the very beginning of this chapter).

The metric as written above is regular and non-degenerate for $2M < r <
\infty$ and $-\infty < t < \infty$.  The singularity at $r=2M$ is not
a true physical singularity \cite{Wald:1984} and can be removed by the
transformation $(t,r) \rightarrow (v,r)$ where
\begin{equation}
 dv = dt + \left(1-\frac{2M}{r}\right) dr\,.
\end{equation}
In these coordinates (the ingoing Eddington-Finkelstein coordinates)
the metric becomes
\begin{equation}
  \label{eq:sch-ief}
  ds^2 = -\left(1-\frac{2M}{r}\right) dv^2 + 2dv\,dr  + r^2d\Omega^2\,,
\end{equation}
where $d\Omega^2 := d\theta^2 + \sin^2\theta d\phi^2$.  The metric is
now regular for $r>0$ and $-\infty < v < \infty$.  We could extend the
solution further by going to double null coordinates, but this shall
suffice for now.

Consider now the vector fields
\begin{equation}
  \label{eq:ell-n-sch}
  \ell = \frac{\partial}{\partial v} + 
  \frac{1}{2}\left(1-\frac{2M}{r}\right)\frac{\partial}{\partial r}\,,\qquad 
  n = -\frac{\partial}{\partial r}\,.
\end{equation}
It is easy to check that these (future directed) vector fields are
both null, i.e. $\ell\cdot\ell = n\cdot n = 0$, and $\ell\cdot n =
-1$.  By convention, we take $\ell$ to be outward pointing and $n$ to
be inward pointing; we have designated $r\rightarrow\infty$ to be
``outwards''.

\begin{figure}
  \centering
  \includegraphics[width=0.8\textwidth]{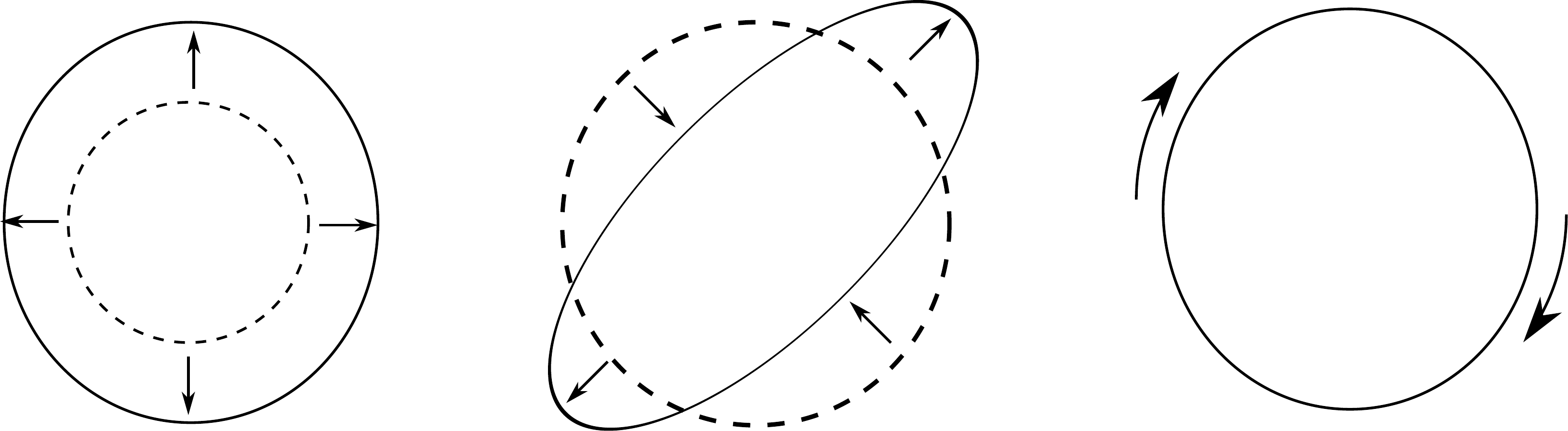}
  \caption{From left to right, an illustration of expansion, shear and
    twist respectively in the $(e_{(1)},e_{(2)})$ plane transverse to
    $\xi^a$, as illustrated by the effect on circles in this plane.
    The expansion $\Theta_{(\xi)}$ is an isotropic expansion, the
    shear is an expansion and contraction in orthogonal directions with
    the area being preserved, and the twist is a rotation.}
  \label{fig:optical}
\end{figure}

Let us pause to recall the notions of \emph{expansion} and
\emph{shear} of a vector field.  For a timelike vector field $\xi$,
the set of vectors orthogonal to $\xi$ form a three-dimensional
plane at each point.  If $A$ is an infinitesimal area element in this
plane and $\lambda$ the affine parameter along $\xi$, then the
expansion of $\xi$ is defined as
\begin{equation}
  \label{eq:expansion-def}
  \Theta_{(\xi)} = \frac{1}{A}\frac{dA}{d\lambda}\,.
\end{equation}
Since a null vector field is orthogonal to itself, it is easy to show
that any vector field $V$ satisfying $V\cdot\xi = 0$ can be written as
a linear combination $V=\alpha\xi + \beta \mathbf{e}_{(1)} +
\gamma\mathbf{e}_{(2)}$ where $\mathbf{e}_{(1)}$ and
$\mathbf{e}_{(2)}$ are mutually orthogonal unit spacelike vectors
orthogonal to $\xi$, and $\alpha,\beta,\gamma$ are real numbers.  The
expansion of $\xi$ is then defined as in Eq.~(\ref{eq:expansion-def})
above except that the relevant area element $A$ is in the
two-dimensional plane spanned by $\mathbf{e}_{(1)}$ and
$\mathbf{e}_{(2)}$.  An alternate expression for the expansion which
is usually more useful is
\begin{equation}
  \Theta_{(\xi)} = q^{ab}\nabla_a\xi_b
\end{equation}
where $q^{ab}$ is the (inverse of) the Riemannian metric in the
$(e_{(1)},e_{(2)})$ plane.  The trace of $\nabla_a\xi_b$ after
projection is the expansion.  The symmetric trace-free part and the
antisymmetric parts give the shear $\sigma_{ab}$ and twist
$\omega_{ab}$ respectively:
\begin{equation}
  \sigma_{ab} = \left(\nabla_{(a}\xi_{b)}\right)^\top - \frac{1}{2}\Theta_{(\xi)}q_{ab}\,,\qquad \omega_{ab} = \left(\nabla_{[a}\xi_{b]}\right)^\top\,.
\end{equation}
where the symbol $(\cdots)^\top$ indicates a projection in the
$(e_{(1)},e_{(2)})$ plane.  The operators $\sigma_{ab}$ and
$\omega_{ab}$ are responsible for transforming vectors in the
$(e_{(1)},e_{(2)})$ plane.  Consider now a set of neighboring null
geodesics generated by $\xi^a$. Let $\zeta^a$ be a connecting vector,
i.e. it is transverse to $\xi^a$ and is Lie dragged along $\xi^a$:
$\Lie_\xi\zeta^a = [\xi,\zeta]^a= \xi^b\nabla_b\zeta^a -
\zeta^b\nabla_b\xi^a= 0$.  We get the effect of the expansion, shear
and twist as operators in the $(e_{(1)},e_{(2)})$ plane leading to the
evolution of $\zeta^a$ in time
\begin{equation}
  \label{eq:zetadot}
  \dot{\zeta}_b^\top := \left(\xi^a\nabla_a\zeta_b\right)^\top = \zeta^a\left(\nabla_a\xi_b\right)^\top 
  = \left(\sigma_{ab} + \omega_{ab} + \frac{1}{2}\Theta_{(\xi)}q_{ab} \right)\zeta^a\,.
\end{equation}
The effect of expansion, shear and twist on $\zeta^a$ is illustrated
in Fig.~\ref{fig:optical}.  A particularly important result is the
Raychaudhuri equation which gives the time derivative of the expansion
for affinely parameterized null-geodesics:
\begin{equation}
  \frac{d \Theta_{(\xi)}}{d\lambda} = -\frac{1}{2}\Theta_{(\xi)}^2 - \sigma_{ab}\sigma^{ab} + \omega_{ab}\omega^{ab} - T_{ab}\xi^a\xi^b\,.
\end{equation}
This is a particular component of the Einstein field equations and a
derivation and applications can be found in
e.g. \cite{HawkingEllis:1973,Wald:1984}.  We shall have occasion to
use it at various points during the course of this chapter.

Returning to the vector fields $\ell^a$ and $n^a$ defined in
Eq.~(\ref{eq:ell-n-sch}), we see that they are manifestly orthogonal
to the constant $(v,r)$ spheres.  Thus their expansions must involve
area elements on these spheres and it is in fact easy to calculate
their expansions:
\begin{equation}
  \label{eq:expansions}
  \Theta_{(\ell)}(r) = \frac{r-2M}{r^2}\,, \qquad \Theta_{(n)}(r) = -\frac{2}{r}\,.
\end{equation}
For spheres outside the black hole region, i.e. spheres with $r>2M$,
we see that $\Theta_{(\ell)} > 0$ and $\Theta_{(n)} < 0$.  This is how
a round sphere in flat space behaves.  However, for spheres in the
black hole region, we get both expansions to be negative.  Such
spheres are known as trapped surfaces and play a fundamental role in
black hole theory and, in particular, in the singularity theorems.
The spheres on the $r=2M$ hyper-surface have $\Theta_{(\ell)} = 0$,
$\Theta_{(n)} < 0$ and are called \emph{marginally} trapped
surfaces. Thus, we see that the $r=2M$ hyper-surface separates the
region where the spherically symmetric trapped surfaces live and are a
signature of a black hole spacetime.  It is worth noting that the
presence of trapped surfaces is not necessarily a signature of strong
field gravity.  The Riemann tensor (and thus the tidal force) is
proportional to $M/r^3$. Thus at $r=2M$, it is $\propto 1/M^2$, and
large black holes have a correspondingly weaker curvature at their
Schwarzschild radius.  In this sense, trapped surfaces are a
non-perturbative phenomena in general relativity.  An observer falling
into a sufficiently large black will not notice anything out of the
ordinary.

\begin{figure}
  \centering
  \includegraphics[width=0.8\textwidth]{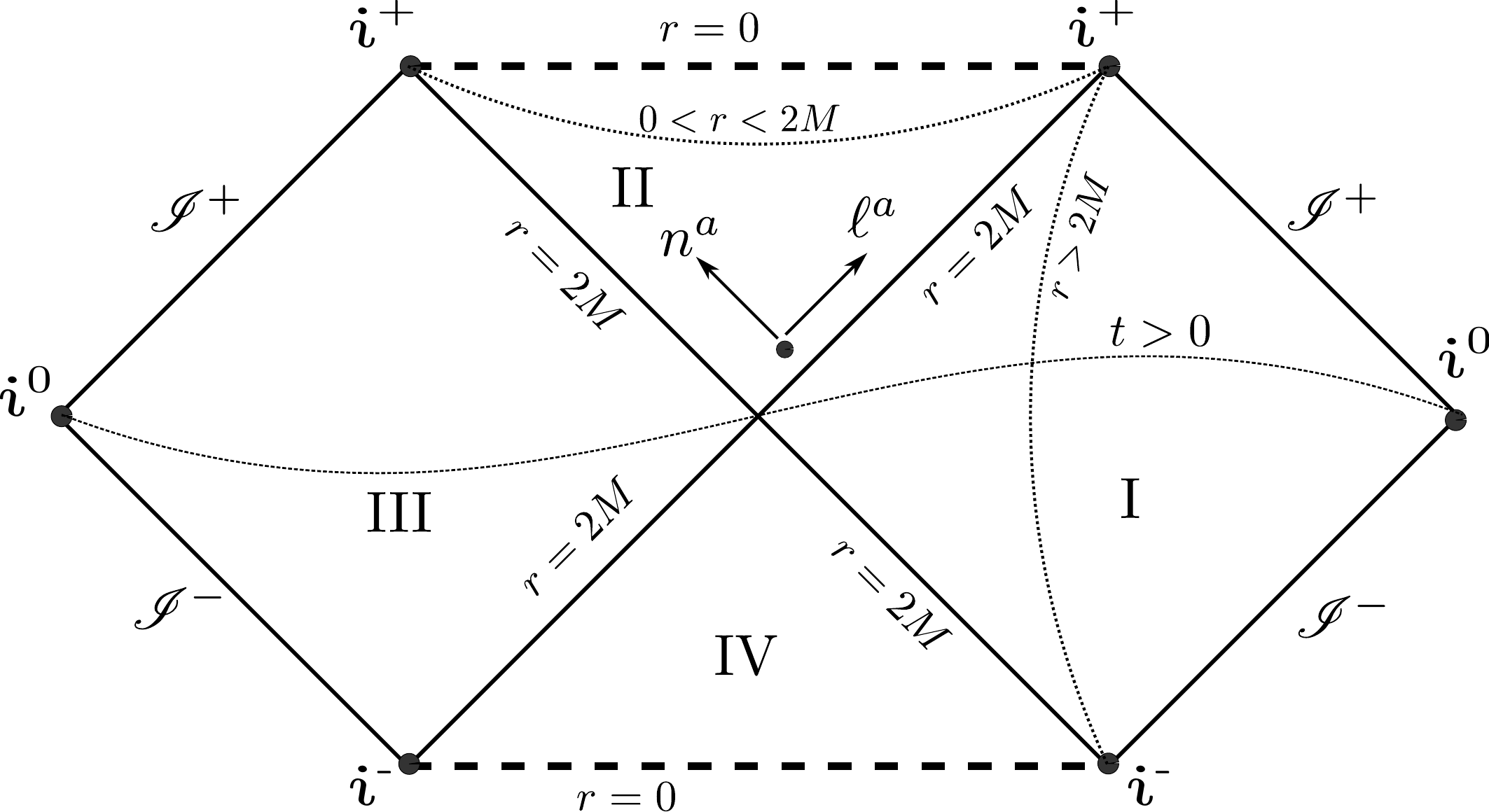}
  \caption{Penrose-Carter conformal diagram for the extended
    Schwarzschild spacetime.  See text for details.}
  \label{fig:schwarzschild}
\end{figure}

The role of the $r=2M$ hyper-surface as the boundary of the region
containing trapped surfaces (known as the trapped region) is typically
not emphasized in standard textbooks on the subject.  What is
emphasized is instead the fact that starting from a point with $r<2M$,
there exist no timelike or null curves which can cross the $r=2M$
hyper-surface.  This is easiest to visualize in a Penrose-Carter
conformal diagram shown in Fig.~\ref{fig:schwarzschild}.  This is a
convenient way of visualizing the $r$-$t$ part of the Schwarzschild
metric.  Just as we had extended the Schwarzschild metric in
Eq.~(\ref{eq:sch-standard}) across $r=2M$ by using ingoing null
coordinates, we can extend the metric of Eq.~(\ref{eq:sch-ief})
further by going to double null coordinates $(u,v)$ where
\begin{equation}
  \label{eq:dbl-null}
  du = dt - \left(1-\frac{2M}{r}\right) dr\,,\qquad dv = dt + \left(1-\frac{2M}{r}\right) dr\,.  
\end{equation}
Fig.~\ref{fig:schwarzschild} is then obtained by performing a further
re-scaling of coordinates and a conformal transformation which brings
infinity to a finite distance.  Details can be found in
\cite{Wald:1984,Stewart:1991} or in other standard textbooks on the
subject.  The original Schwarzschild metric is region I in this
diagram while Eq.~(\ref{eq:sch-ief}) corresponds to I and II.  Regions
III and IV are mirror images of I and II respectively.  In this
figure, null curves are straight lines at $45^\circ$, $i^0$ is spatial
infinity, $i^+$ is future timelike infinity and $i^-$ is past timelike
infinity.  Future directed null curves in this figure end up either at
the future singularity at $r=0$ (marked with a dashed line) or at
future null-infinity labeled as $\mathscr{I}^+$ ($\mathscr{I}^-$ is
past null-infinity).  It is also worth pointing out that if region I
is defined to be the ``outside world'' so that $\ell^a$ is the outward
pointing null normal (this is merely a matter of convention), then
$\Theta_{(\ell)}<0$ in regions II \emph{and} III.  However, region III
has $\Theta_{(n)}>0$ so that only region II has both expansions
negative.  Similarly, only region IV has both expansions positive.

It is then clear that no point in region II can be connected with
region I (or III) by a causal curve and thus justifies the term
``black hole'' for region II.  The boundary of this region occurs at
$r=2M$ and is called an event horizon.  Note that the boundary of
region IV also occurs at $r=2M$ but region IV is a time-reversed
version of II; it has the property that no future directed causal
curve can stay within it.  Thus, region IV is called a white-hole.
In the rest of this chapter, we shall only consider the portion of the
$r=2M$ hyper-surface which bounds the black hole, i.e. region II.
In a physical gravitational collapse situation, the white-hole (and
region III) does not actually exist and is covered up by the matter
fields which constitute the star; this will soon be seen explicitly
when we study the Vaidya metric.

We thus have two different routes for describing a black hole: trapped
surfaces versus event horizons (though one might suspect
\emph{a-priori} that the two might be intimately related).  Trapped
surfaces seem to be more local than event horizons. To know that a
particular null geodesic will not leave a particular region of
spacetime, one might need to know properties of the spacetime far away
from the starting point of the geodesic.  On the other hand, for
trapped surfaces, in the Schwarzschild case we have just needed the
computations of the expansions in Eq.~(\ref{eq:expansions}) at a fixed
value of $(v,r)$.  However, one shouldn't forget that the computations
of the expansions are not at just a single spacetime point, but are
instead to be performed at all points over a sphere. This issue is
irrelevant for spherically symmetric trapped surfaces in spherically
symmetric spacetimes, but it is an important point that one cannot
identify a trapped surface by examining only a part of it.  For this
reason, the trapped surface condition is said to be
\emph{quasi-local}.  In any case, for Schwarzschild, the two
descriptions of the black hole region agree: the $r=2M$ hyper-surface is
both the event horizon and also the boundary of the region of
spacetime which contains trapped surfaces.

\begin{figure}
  \centering
  \includegraphics[width=0.8\textwidth]{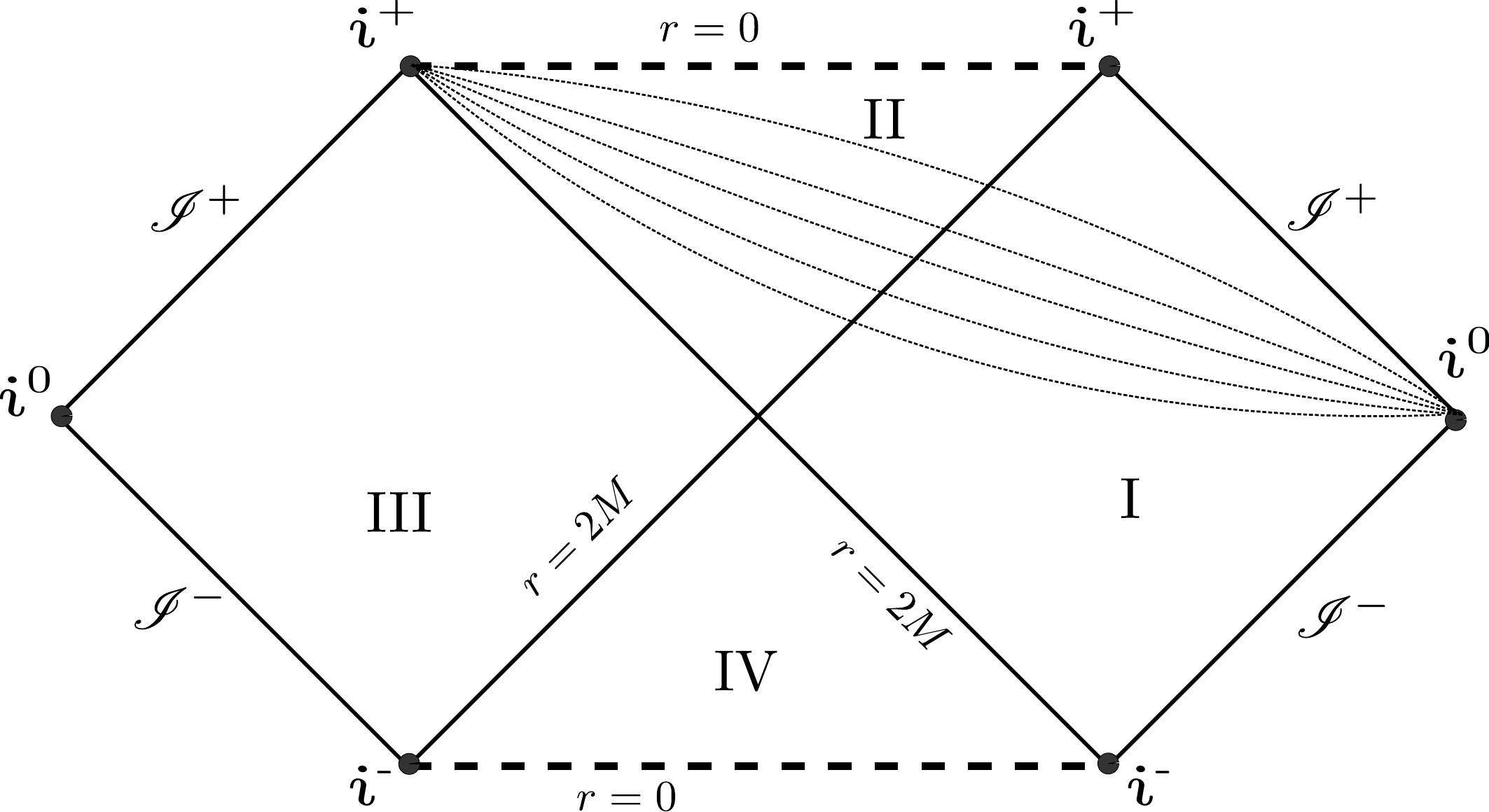}
  \caption{A non-spherically symmetric spatial hyper-surface in the
    Schwarzschild spacetime depicted as a set of curves.}
  \label{fig:schwarzschild2}
\end{figure}

Visualizing non-spherically symmetric trapped surfaces is harder, even
in a spacetime as simple as Schwarzschild.  As in many numerical
investigations in general relativity, let us try to locate such
surfaces on three-dimensional spatial hyper-surfaces.  The equation
$\Theta_{(\ell)}=0$ turns into a minimization problem, and to a
second-order elliptic equation in three dimensional space (we shall
have more to say on this matter later).  The spatial hyper-surfaces
depicted in Fig.~\ref{fig:schwarzschild} were all spherically
symmetric, and thus a single curve in the Penrose diagram suffices for
them.  However, if we wish to depict non-spherically symmetric
hyper-surfaces, we will need a collection of such curves, say one for
each value of $(\theta,\phi)$.  An example is shown in
Fig.~\ref{fig:schwarzschild2}.  In this example, the spatial
hyper-surface starts from $i^+$ on region III (but this detail is not
important for our purposes and we could as well have started from
$i^0$ on the left edge of region III).  What is important is that the
spatial hyper-surfaces, or alternatively all the curves shown in
Fig.~\ref{fig:schwarzschild2} intersect the event horizon.  As we
shall see later, due to the somewhat non-intuitive properties of such
null surfaces, the intersection of such a spatial hyper-surface with
the $r=2M$ hyper-surface is still a marginally trapped surface though
now a non-symmetric one.  More specifically, it turns out that
$\ell_a$ is covariantly constant on the $r=2M$ surface so that
$\nabla_a \ell_b$, projected onto the $r=2M$ surface, vanishes
identically (see the discussion around Eq.~(\ref{eq:xdy0}) and in
Sec.~\ref{subsec:ihdefs}).  This means that all closed cross-sections
of the $r=2M$ surface are marginally trapped.

There would of course generally be non-spherically symmetric trapped
surfaces on these spatial hyper-surfaces lying inside the marginally
trapped one.  Each spherically symmetric trapped and marginally
trapped surface can be found by such a procedure.  This construction
clearly shows that there are many more non-symmetric trapped surfaces
than symmetric ones; each spherically symmetric hyper-surface can be
deformed in an infinite number of ways and still contain trapped and
marginally trapped surfaces.

We note finally that it is possible to come up with examples where
part of the spatial hyper-surface extends arbitrarily close to the
future singularity, but part of it is still outside the black hole
region so that its intersection with the event horizon is not a
complete sphere.  There would then exist no marginally trapped
surfaces on such a spatial hyper-surface \cite{Wald:1991zz}.

\subsection{The Vaidya spacetime}
\label{subsec:vaidya}

As we have just seen, for a Schwarzschild black hole, all the natural
definitions of the surface of a black hole agree.  Thus, the $r=2M$
surface is both the boundary of the trapped region and also the event
horizon.  This has led to a widespread belief that the notion of a
black hole and its surface is unambiguous.  Matters are however not so
simple in dynamical situations.  Let us now look at what is perhaps
the simplest example of a dynamical black hole, namely the spherically
symmetric Vaidya spacetime \cite{Vaidya:1951zz}.

\begin{figure}
  \centering
  \includegraphics[width=0.4\textwidth]{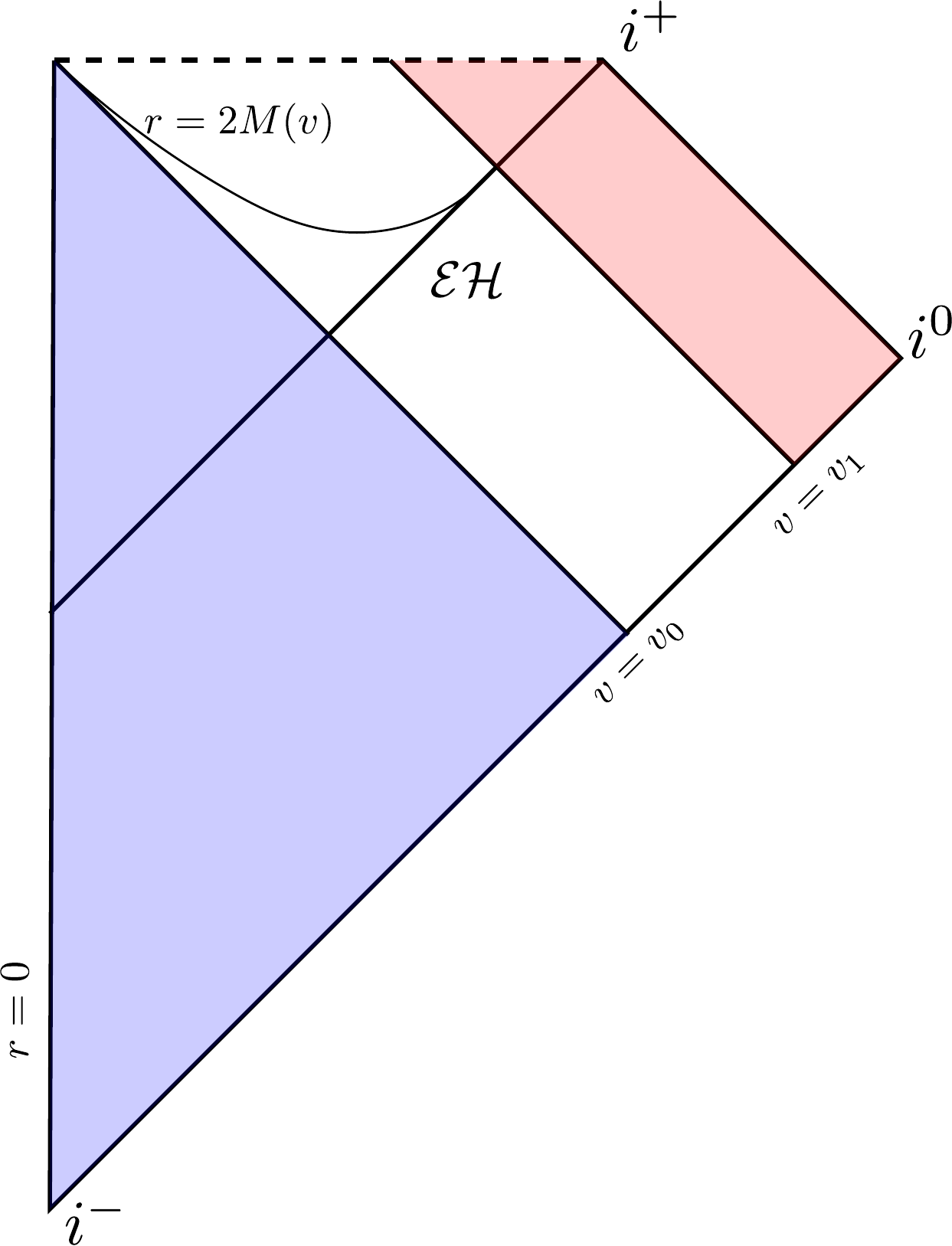}
  \caption{Penrose-Carter conformal diagram for the Vaidya spacetime.
    The region shaded in blue is flat and the region in red is
    isomorphic to a portion of Schwarzschild.  The event horizon is
    labeled $\mathcal{EH}$ and is seen to be distinct from the
    $r=2M(v)$ surface.  The two agree only in the final Schwarzschild
    portion.}
  \label{fig:vaidya}
\end{figure}

The Vaidya solution is obtained by starting with the Schwarzschild
metric in ingoing Eddington-Finkelstein coordinates of
Eq.~(\ref{eq:sch-ief}), and replacing the constant $M$ by a
non-decreasing function $M(v)$:
\begin{equation}
  \label{eq:vaidya}
  ds^2 = -\left(1-\frac{2M(v)}{r}\right) dv^2 + 2dv\,dr  + r^2d\Omega^2\,.
\end{equation}
The stress energy tensor for this metric is 
\begin{equation}
  \label{eq:vaidya-Tab}
  T_{ab} = \frac{\dot{M}(v)}{4\pi r^2} \nabla_a v \nabla_bv\,,\qquad \textrm{where} \qquad \dot{M}(v) := \frac{dM(v)}{dv}\,.
\end{equation}
This represents the collapse of null-dust to form a black hole and if
$\dot{M}(v)\geq 0$ then $T_{ab}$ satisfies the dominant energy
condition.  If we choose a mass-function which is non-zero only for
$v>v_0$ and constant after $v=v_1$, then we will have (portions of)
Minkowski and Schwarzschild spacetimes for $v<v_0$ and $v>v_1$
respectively.  The Penrose-Carter diagram for this spacetime is shown
in Fig.~\ref{fig:vaidya}.  A suitable set of null normals orthogonal
to the constant $(r,v)$ spheres are
\begin{equation}
  \label{eq:ell-n-vaidya}
  \ell = \frac{\partial}{\partial v} + 
  \frac{1}{2}\left(1-\frac{2M(v)}{r}\right)\frac{\partial}{\partial r}\,,\qquad 
  n = -\frac{\partial}{\partial r}\,,
\end{equation}
and their expansions are respectively found to be
\begin{equation}
  \label{eq:expansions-vaidya}
  \Theta_{(\ell)}(v,r) = \frac{r-2M(v)}{r^2}\,, \qquad \Theta_{(n)}(v,r) = -\frac{2}{r}\,.
\end{equation}
Thus, in this case, there are no spherically symmetric trapped
surfaces outside the $r=2M(v)$ surfaces and as in Schwarzschild, the
spheres with $r=2M(v)$ and fixed $v$ are marginally trapped surfaces.

The event horizon is also not difficult to locate. Consider the
outgoing null geodesics generated by the vector field $\ell$ above.
Some of these geodesics will reach infinity while others will
terminate at the singularity.  The event horizon is the boundary
between the two cases.  If we assume that the mass function reaches a
finite final steady state value $M_\infty$, then the final black hole
is a portion of Schwarzschild, and thus the condition $r=2M_\infty$
defines the final state of the event horizon.  Thus, we want to find
the outgoing null geodesic generated by $\ell^a$ defined in
Eq.~(\ref{eq:ell-n-vaidya}) for which $r\rightarrow 2M_\infty$ when
$v\rightarrow \infty$.  Subject to this final state boundary
condition, we need to solve
\begin{equation}
  \frac{dr}{dv} = \frac{1}{2}\left( 1-\frac{2M(v)}{r} \right)\,.
\end{equation}
This is fairly easy to solve numerically for a generic mass function.

\begin{figure}
  \centering
  \includegraphics[width=0.4\textwidth]{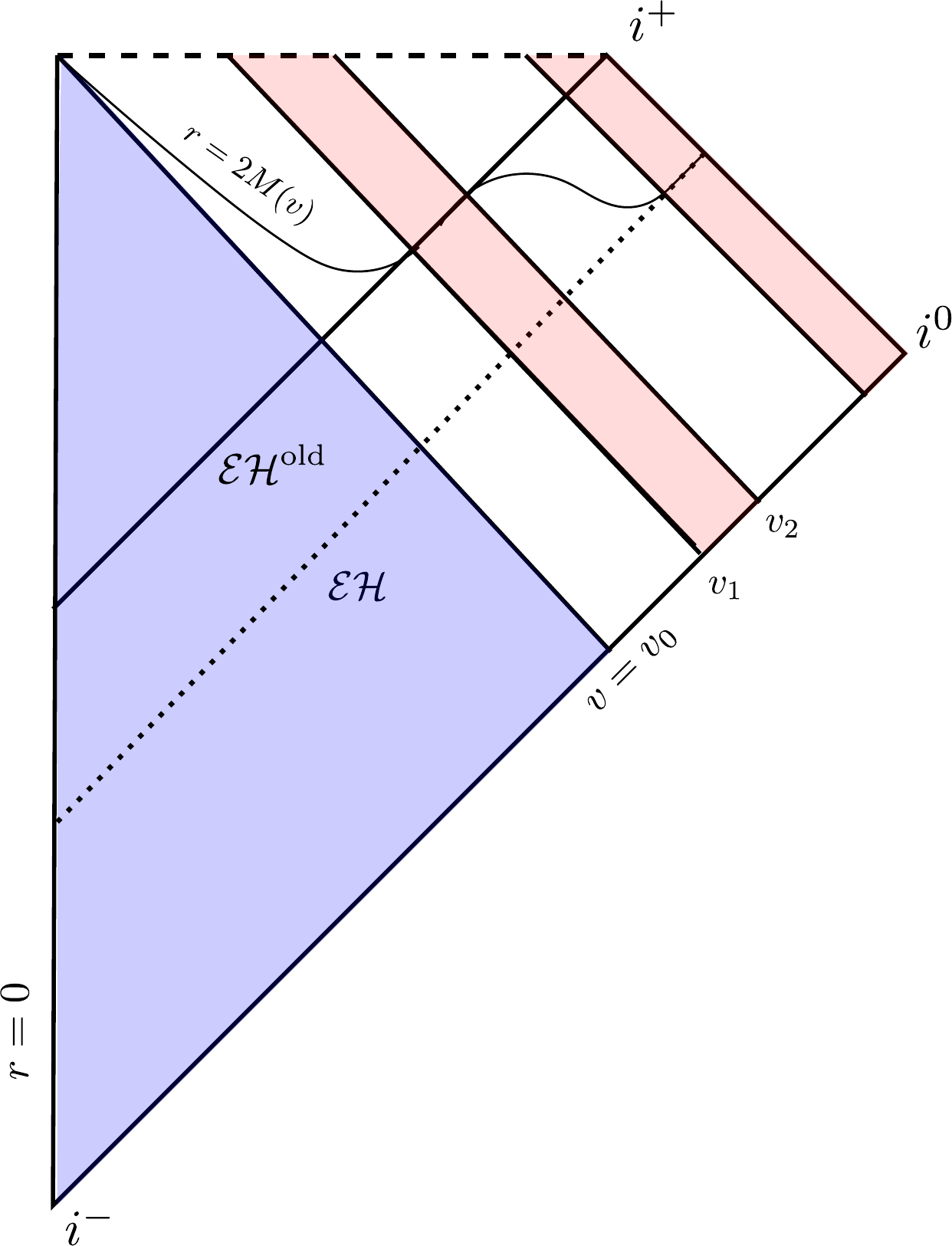}
  \caption{Another Penrose-Carter conformal diagram for the Vaidya
    spacetime.  The region shaded in blue is again the flat
    region. The mass function here has two phases where it is
    increasing.  An observer with $v<v_2$ could in principle locate
    the solid line marked $\mathcal{EH}^{\rm old}$ but will not know
    that the mass function will increase again and that the true event
    horizon is in fact given by the surface (indicated by a dotted
    line in this figure) marked $\mathcal{EH}$.}
  \label{fig:vaidya2}
\end{figure}

Let us now note a few properties of this event horizon. A typical case
is shown in Fig.~\ref{fig:vaidya}. First note that the event horizon
extends to the flat region.  A mortal observer in the flat region who
has no way of knowing that the gravitational collapse will occur at
some time in the future, might actually be living near an event
horizon.  The existence of the event horizon has really no
consequences for any physical experiment or observations that the
observer can conduct locally, and contrary to popular belief, the
observer can cross the event horizon without feeling anything out of
the ordinary.  Furthermore, even an observer in the intermediate
region $v_0<v<v_1$, who can witness gravitational collapse occurring
cannot know the true location of the event horizon.  To illustrate
this, consider Fig.~\ref{fig:vaidya2}.  Here the mass function is
non-zero for $v<v_0$ as before, however, there are two phases.  The
mass function first reaches a constant value at $v_1$ but restarts
again at a later time $v_2$.  The observer with $v<v_2$ cannot know
the value of $M_\infty$ and can thus never know the true location of
the event horizon.

\subsubsection{Examples of non-symmetric trapped surfaces}

In contrast to the event horizon, the $r=2M(v)$ surface seems to have
the right properties. It bounds the region which has spherically
symmetric trapped surfaces, it doesn't extend into the flat region, it
grows only when $\dot{M}>0$, it can be located quasi-locally,
i.e. by checking the conditions for a trapped surface on a sphere, and
it doesn't care about what happens to $M(v)$ at late times. However,
non-spherically symmetric trapped surfaces are not so well behaved.
While there are no marginally trapped surfaces which lie completely
within the flat region, we shall see that portions of them can extend
into the flat region.

\begin{figure}
  \centering
  \includegraphics[width=0.7\textwidth]{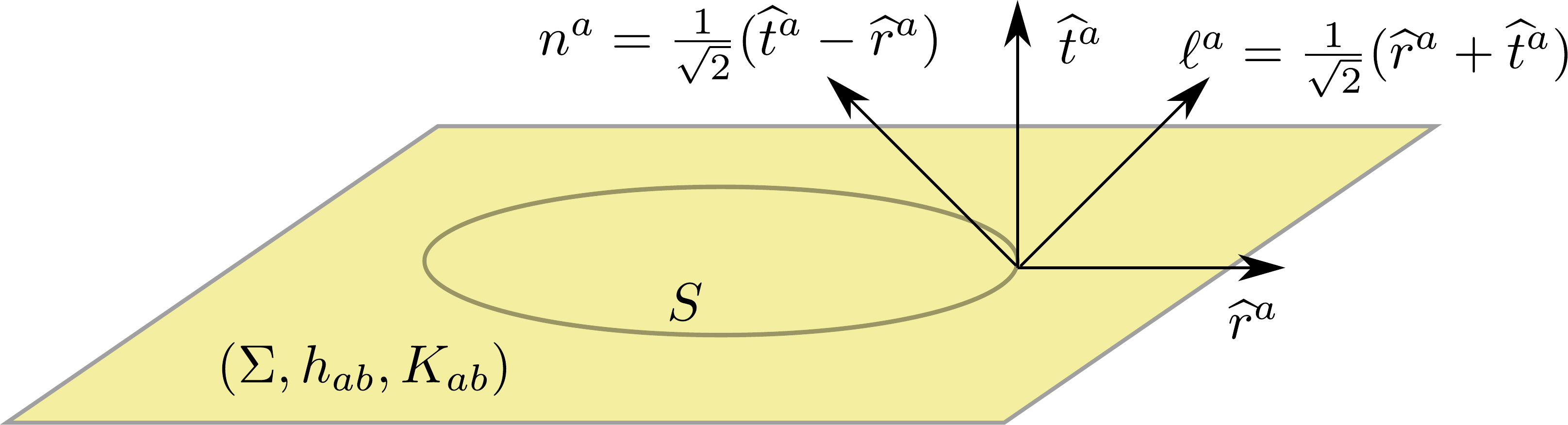}
  \caption{A closed marginally outer trapped surface on a spatial
    Cauchy surface $\Sigma$ with intrinsic metric $h_{ab}$ and
    extrinsic curvature $K_{ab}$.  The unit timelike normal to
    $\Sigma$ is $\widehat{t}^a$ and the outward unit spatial normal to
    $S$ is $\widehat{r}^a$.  A particular choice of the out- and
    in-going null normals are $\frac{1}{\sqrt{2}}(\widehat{t}^a\pm
    \widehat{r}^a)$. }
  \label{fig:ah}
\end{figure}

\begin{figure}
  \centering
  \includegraphics[width=0.7\textwidth]{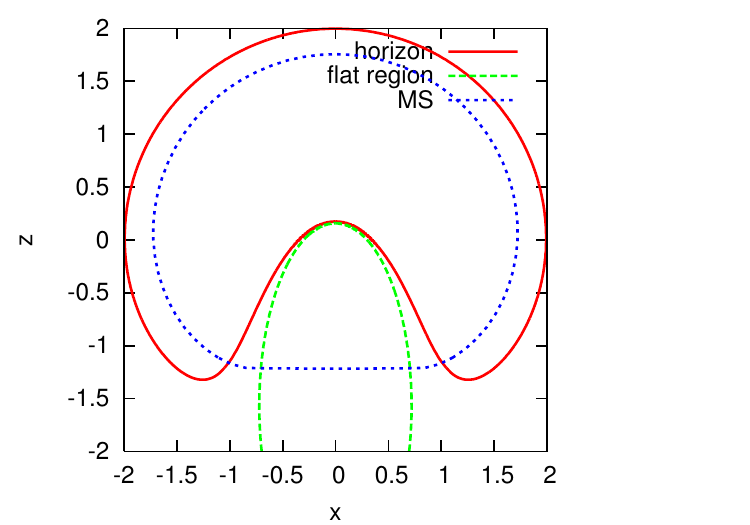}
  \caption{A non-symmetric closed surface in a particular Vaidya
    spacetime with $\Theta_{(\ell)}=0$ and $\Theta_{(n)}\leq 0$. The
    solid red curve is the intersection of the spatial hyper-surface with the
    $r=2M(v)$ surface, the green dotted line is the boundary of the
    flat region, and the blue dashed line is the marginal surface
    located on this spatial hyper-surface. }
  \label{fig:vaidya-numrel}
\end{figure}

One can try to find marginally trapped surfaces numerically.  The
standard procedure is to start with a particular spatial hyper-surface
$\Sigma$, and to find a surface $S$ in $\Sigma$ for which
$\Theta_{(\ell)} = 0$.  Let $h_{ab}$ be the Riemannian metric on
$\Sigma$ induced by $g_{ab}$. If the unit normal on $\Sigma$ is
$\widehat{t}^a$, the unit spacelike normal to $S$ on $\Sigma$ is
$\widehat{r}^a$ with $\ell^a = (\widehat{t}^a +
\widehat{r}^a)/\sqrt{2}$ (see Fig.~\ref{fig:ah}). With this choice,
noting that the metric on $S$ is $q_{ab} = h_{ab} -
\widehat{r}_a\widehat{r}_b = g_{ab} + \widehat{t}_a\widehat{t}_b -
\widehat{r}_a\widehat{r}_b$, the condition $\Theta_{(\ell)}=0$ can be
written as
\begin{equation}
  \sqrt{2}\Theta_{(\ell)} = \sqrt{2}q^{ab}\nabla_{a}\ell_b = D_a\widehat{r}^a + K_{ab}\widehat{r}^a\widehat{r}^b - K = 0\,.
\end{equation}
Here $K_{ab} = h_{a}^ch_b^d\nabla_c \widehat{t}_d$ is the extrinsic
curvature, $K$ is its trace, and $D$ is the derivative operator on
$\Sigma$ (we shall explain these concepts is more detail later in
Sec.~\ref{subsec:trapped}). Taking coordinates $(r,\theta,\phi)$ on
$S$ and assuming that the surface is given by an equation
$r=f(\theta,\phi)$, the above equation becomes a non-linear
second-order partial differential equation for $f$ which can be solved
numerically.  Typical methods assume that the surface is
\emph{star-shaped}, i.e. every ray from the origin $r=0$ intersects
the surface exactly once; for a more complete description of this and
other methods, we refer to \cite{Thornburg:2006zb}.
We then choose a particular mass function and a $\Sigma$ defined
through a particular non-axisymmetric time coordinate and attempt to
locate surfaces with $\Theta_{(\ell)} = 0$.  Let us review an
illustrative result from a study reported in \cite{Schnetter:2005ea}
(see also \cite{Nielsen:2010wq} for another such study) The particular
mass function chosen corresponds to a short pulse of radiation:
\begin{equation}
  M(v) =  \begin{array}{l}
      0 \quad \textrm{for} \quad v\leq 0\,, \\
      M_0v^2/(v^2+W^2) \quad \textrm{for} \quad v> 0\,. 
    \end{array} 
\end{equation}
The parameter $M_0$ is the final mass and $W$ determines the
time-scale of the radiation pulse.  We choose $M_0=1$ and $W=0.1$.
The non-spherically symmetric time coordinate is taken to be
\begin{equation}
  \bar{t} = v - r(1+\alpha\cos\theta)\,.
\end{equation}
The constant $\alpha$ determines the degree of asymmetry, and we
choose $\alpha=10/11$.  As in Fig.~\ref{fig:schwarzschild2}, spatial
hyper-surfaces of constant $\bar{t}$ correspond to different sets of
curves (one for each $\theta$) in the Penrose-Carter diagram.
Fig.~\ref{fig:vaidya-numrel} shows the marginal surface found on the
$\bar{t}=-0.3$ spatial hyper-surface.  It shows the section in the
$x-z$ plane.  The blue dotted line is the marginal surface, the green
dashed line encloses the intersection of the flat region with the
hyper-surface, and the solid red curve shows the intersection with the
$r=2M(v)$ surface.  $\bar{t}=-0.3$ surface (which is not a marginal
surface).  Thus, we see that the marginal surface extends in the flat
region, though in this example it is planar with $\Theta_{(n)} = 0$
(so it is not, strictly speaking, a marginally trapped surface).  The
marginal surface is seen to be only partially inside the $r=2M(v)$
surface.

\subsubsection{The trapped region}

Having looked at particular examples of trapped surfaces in a Vaidya
spacetime, let us consider the trapped region, i.e. the portion of the
manifold which contains trapped surfaces.  The trapped region for
Vaidya spacetime can in fact be studied analytically.  The starting
point for this goes back to a conjecture by Eardley in 1998
\cite{Eardley:1997hk}: \emph{The outer boundary of the region
  containing outer trapped surfaces is the event horizon} (an outer
trapped surface has $\Theta_{(\ell)}<0 $ and no restriction on
$\Theta_{(n)}$).  One the one hand, it is known that trapped surfaces
cannot cross the event horizon.  On the other hand, in dynamical
situations like Vaidya, the event horizon is growing in area and its
cross-sections are not marginally trapped.  Thus, while the outer
trapped surface might get arbitrarily close to the event horizon, the
limiting process is not trivial.  The Vaidya spacetime provides a
relatively simple setting to study this phenomena.

Recent works in this direction have been
\cite{Schnetter:2005ea,BenDov:2006vw,Bengtsson:2008jr,Bengtsson:2010tj}).
For any point $p$ in the flat portion of the black hole region of the
Vaidya spacetime, Ben-Dov showed \cite{BenDov:2006vw} that there
exists an outer trapped surface $S$ which contains $p$.  This works
even when $p$ is arbitrarily close to the event horizon.  In this
case, most of the trapped surface actually lies inside the $r=2M(v)$
surface in the far future where $v$ is large.  There is a narrow
``tendril'' which is almost null for a large portion, and yet
stretches from the far future right down to the flat portion within
the event horizon.  This is precisely the kind of trapped surface
whose existence was conjectured by Eardley in \cite{Eardley:1997hk}.
It is not clear that such highly non-symmetric trapped surfaces would
be present in typical spatial hyper-surfaces used in numerical
relativity simulations, or in fact, whether the standard numerical
methods currently employed would be able to locate such a surface even
if it were present.

It is also possible to locate the boundary of closed future trapped
surfaces (i.e. surfaces with $\Theta_{(\ell)}<0$ and $\Theta_{(n)}<0$)
in Vaidya spacetimes. The first result was obtained by Ben-Dov
\cite{BenDov:2006vw}, but the complete solution to the problem was
found by Bengtsson \& Senovilla
\cite{Bengtsson:2008jr,Bengtsson:2010tj}.  Bengtsson \& Senovilla have
proved a number of results regarding the properties of trapped
surfaces in spherically symmetric spacetimes, but here we shall only
illustrate them by describing the past spacelike barrier for trapped
surfaces in the Vaidya case.  To this end, we need the following
result (Theorem 4.1 of \cite{Bengtsson:2010tj}): In a region
$\mathcal{R}$ of a spacetime, let $\xi^a$ be a future pointing
hyper-surface orthogonal vector field so that $\xi_a =
-F\nabla_a\bar{t}$ for some $F>0$ and some $\bar{t}$ which increases
to the future.  If $S$ is a future trapped surface which intersects
$\mathcal{R}$ (but is not necessarily contained within $\mathcal{R}$),
then $S$ cannot contain a local minimum of $\bar{t}$ at points with
$q^{ab}\Lie_\xi g_{ab} \geq 0$. 

\begin{figure}
  \centering 
  \subfloat[]{
    \label{fig:vaidya3}
    \includegraphics[width=0.4\textwidth]{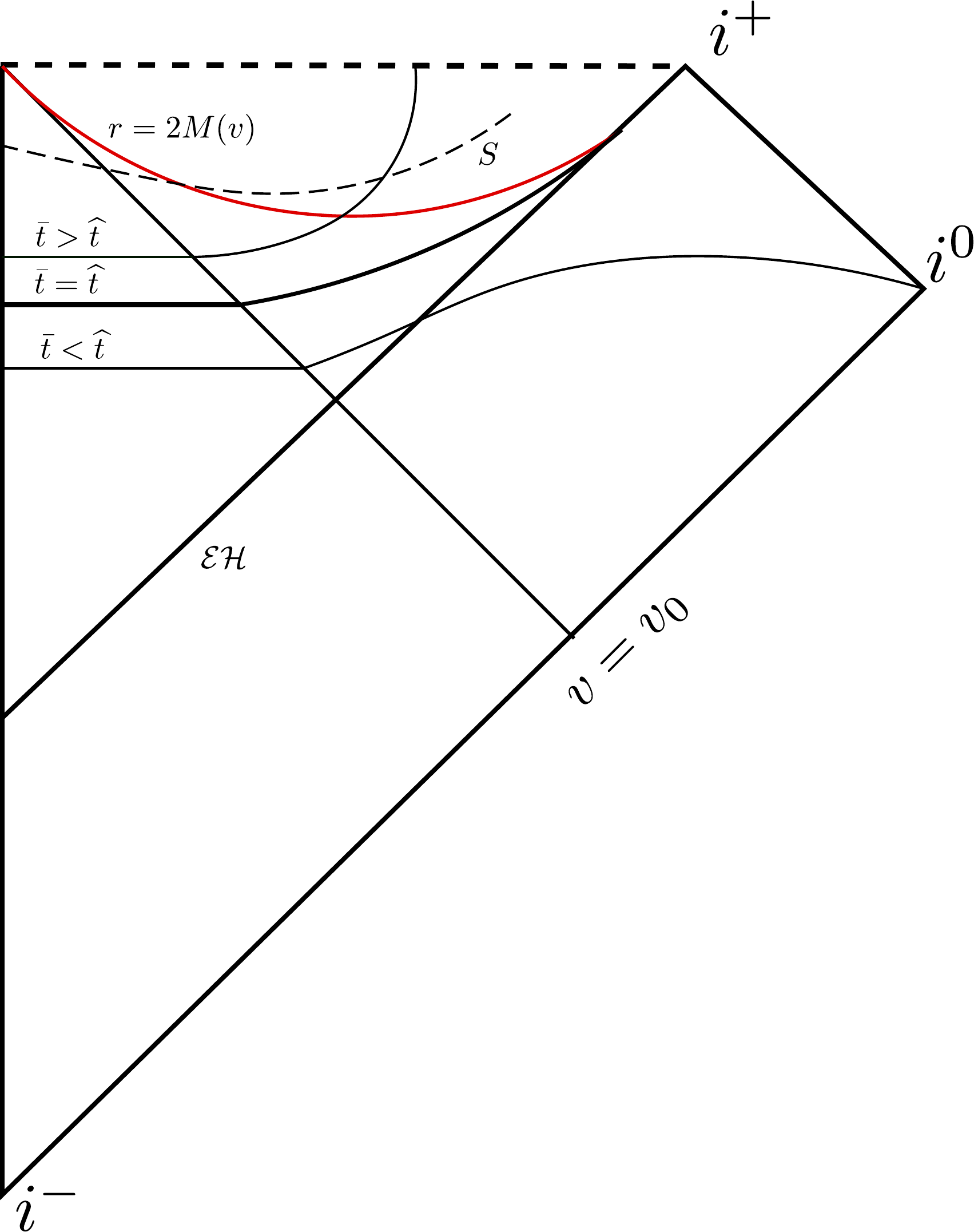}}
  \subfloat[]{
    \label{fig:vaidya3a}
    \includegraphics[width=0.4\textwidth]{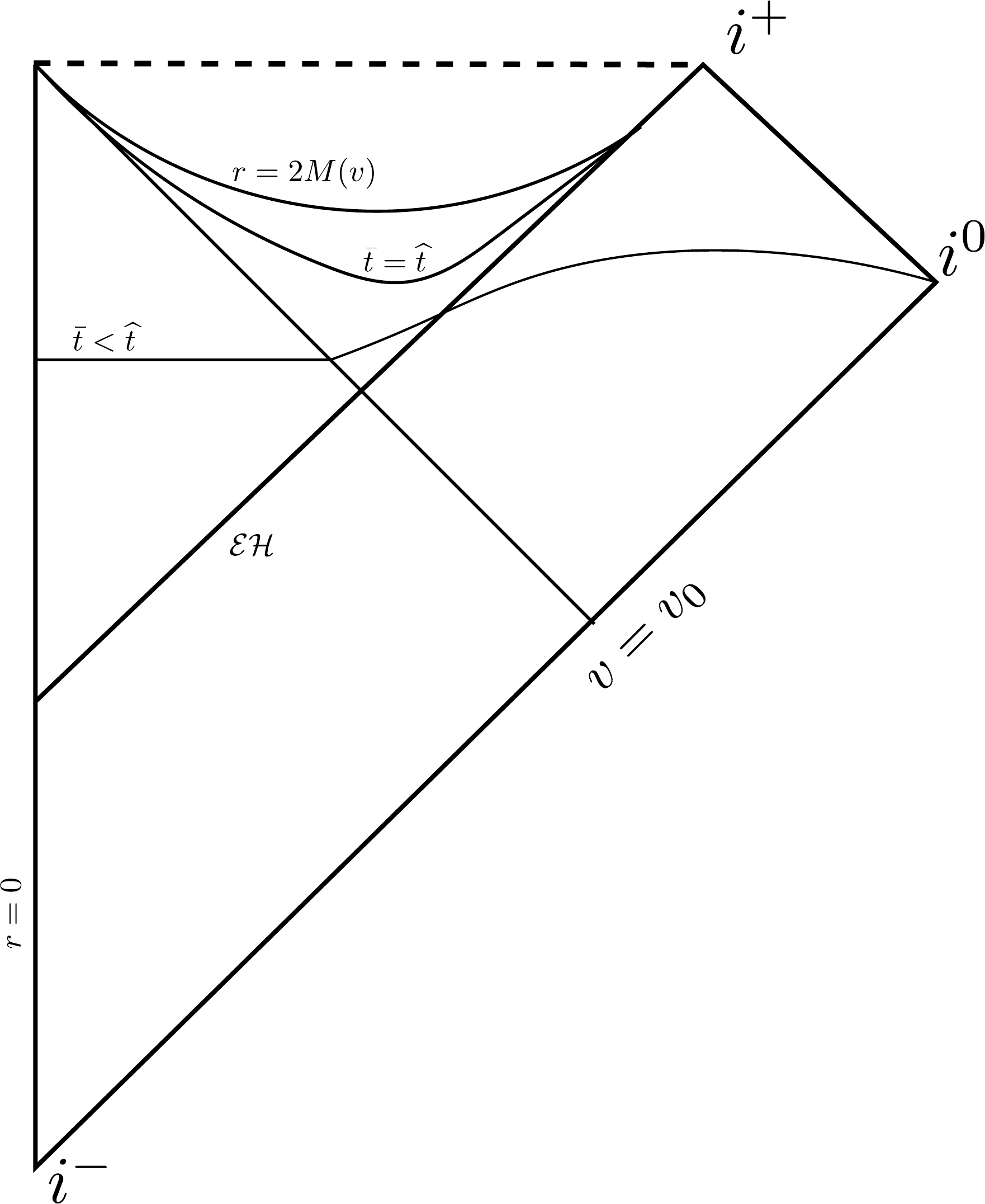}}
  \caption{Surfaces of constant Kodama time for a Vaidya spacetime
    with $\mu:= \lim_{v\rightarrow 0}M(v)/v > 1/8$.  In this case
    $\widehat{\Sigma}$ lies partially in the flat region.  A future
    trapped surface which extends in the flat region is depicted as
    $S$.  It required to enter the $r>2M(v)$ region with increasing
    $\bar{t}$.  The right panel is similar to the left panel, but in
    this case $\widehat{\Sigma}$ does not extend into the flat
    region. This happens for $\mu\leq 8$.  A similar picture also
    works when $M(v)$ asymptotes to a finite value for
    $v\rightarrow\infty$. This figure is essentially Fig.~15 of
    \cite{Bengtsson:2010tj}.}
\end{figure}

In a region $\mathcal{R}$ with a time coordinate such as $\bar{t}$,
the significance of this results is that once a future trapped surface
enters such a region with initially decreasing $\bar{t}$ (i.e. if the
surface is initially ``bending downwards'' in time), then $\bar{t}$
must continue decreasing.  If $\mathcal{R}$ is bounded in the past by
the event horizon, then clearly this result forces $S$ to continue
till it reaches the event horizon.  Since $S$ cannot cross the event
horizon (or even touch it), it becomes clear that the region
$\mathcal{R}$ cannot contain even portions of future trapped surfaces
which are bending downwards in time.  In Vaidya, an appropriate $\xi$
is the so-called Kodama vector
\begin{equation}
  \label{eq:kodama}
  \xi^a = \left(\frac{\partial}{\partial v} \right)^a \implies \xi_a 
  = \nabla_ar - \left(1-\frac{2M(v)}{r} \right)\nabla_av \,.
\end{equation}
Since $\xi_a\xi^a = -(1-2M(v)/r)$ it is clear that $\xi^a$ is future
directed to the past of the $r=2M(v)$ surface.  Furthermore, it is
easy to check that
\begin{equation}
  \Lie_\xi g_{ab} = 2\dot{M}(v)\ell_a\ell_b\,.
\end{equation}
Thus, $q^{ab}\Lie_\xi g_{ab} \geq 0$ if $\dot{M}\geq 0$.  The surfaces
of constant $\bar{t}$, denoted by $\Sigma^{\bar{t}}$ are spherically
symmetric and defined by
\begin{equation}
  \frac{dv}{dr} = \left(1-\frac{2M(v)}{r} \right)^{-1}\,.
\end{equation}
If $M(v)$ is constant for $v>v_1$, then there is a value of
$\bar{t}=\widehat{t}$ such that $\Sigma^{\bar{t}}$ coincides with the
event horizon for $v>v_1$.  Alternatively, if $M(v)$ asymptotes to a
constant value, then there is a value of $\bar{t}=\widehat{t}$ such
that $\Sigma^{\bar{t}}$ also asymptotes to the event horizon. In both
these cases, let us denote these $\Sigma^{\bar{t}}$ as
$\widehat{\Sigma}$.  In the flat region, $\Sigma^{\bar{t}}$ are
horizontal lines in the Penrose diagrams.  The behavior of
$\widehat{\Sigma}$ depends sensitively on $M(v)$.  In particular, it
depends on the quantity $\mu := \lim_{v\rightarrow 0} M(v)/v$.  When
$\mu>1/8$, it is shown in \cite{Bengtsson:2010tj} that
$\widehat{\Sigma}$ will enter the flat region, but in other cases it
may not. These cases are depicted in Figs.~\ref{fig:vaidya3} and
\ref{fig:vaidya3a} along with $\Sigma^{\bar{t}}$ for some other values
of $\bar{t}$.  Any trapped surface which extends outside the $r=2M(v)$
surface must do so with increasing $\bar{t}$ and it must never
``bend'' downwards in time.  If it does not do so, then the above
result shows that it must continue downwards till it hits the event
horizon where it must cease to be smooth if the condition
$\Theta_{(\ell)}<0$ condition is to be maintained (the event horizon
is expanding and thus has positive expansion).  It cannot terminate
smoothly because this would imply the existence of a point on the
trapped surface where $\bar{t}$ is a local minimum.

A little thought then shows that $\widehat{\Sigma}$ must then be a
past barrier which future trapped surfaces cannot cross.  Furthermore,
it is also clear that there cannot be any compact future trapped
surface contained entirely in the region to the past of
$\widehat{\Sigma}$, and bounded between $\widehat{\Sigma}$ and the
event horizon; if there were, again there would have to be a point on
the trapped surface where $\bar{t}$ is a local minimum.  The boundary
of the region containing trapped surfaces has thus been located.  This
boundary $\widehat{\Sigma}$ is of course spherically symmetric (as one
can prove on general grounds).  However, it is not foliated by
marginally trapped surfaces.  Thus, as with the limit of outer trapped
surfaces to the event horizon, it is clear that the the limit of
marginally trapped surfaces to this boundary cannot be smooth.
$\widehat{\Sigma}$ is not a quasi-local object; it is as non-local as
the horizon. Moreover, as far as we know, it does not have any
features which might distinguish it as a black hole horizon.

\subsubsection{Lessons from spherical symmetry}

After this extensive discussion of the Schwarzschild and the Vaidya
examples let us summarize the situation in spherical symmetry.  We
have seen the complications that can arise from having to consider
non-spherically symmetric trapped surfaces in dynamical situations.
In Schwarzschild there are no major surprises and any trapped surface
can extend right up to the event horizon.  This is not so in Vaidya;
the obvious generalization of the $r=2M$ surface from Schwarzschild
does not coincide with the event horizon.  We need non-spherically
symmetric trapped surfaces to ``fill the gap'' between the $r=2M(v)$
surface and the event horizon.  It however turns out to be possible to
study the trapped region in detail and to obtain a fairly complete
understanding of where trapped surfaces can (and cannot) occur.  There
turns out to be a difference between future- and outer-trapped
surfaces (i.e. whether or not we consider the $\Theta_{(n)} < 0$
condition).  Outer trapped surfaces can extend all the way to the
event horizon, but understanding how such surfaces limit to the event
horizon is subtle; there is a separate past barrier for future trapped
surfaces which is distinct from the event horizon.

What does this study tell us about non-spherically symmetric dynamical
spacetimes?  We note that the essential complication here, as noted by
Eardley \cite{Eardley:1997hk}, is not that the black hole is
non-spinning etc. Rather, the problem is to consider trapped surfaces
which do not share the symmetry of the spacetime and to understand
what happens to them as they are deformed towards the event horizon
which will generally have positive expansion in dynamical situations.
We might still expect outer trapped surfaces to extend to the event
horizon in a similar fashion but there is, in general, probably no
separate barrier for future trapped surfaces.

The next step in this chapter will be to study how marginally trapped
surfaces evolve in time and under general deformations, and this leads
to the subject of quasi-local horizons.  However, before doing so, we
shall first formalize many of the ideas introduced in this section
with some general definitions and results.

\section{General definitions and results: Trapped surfaces, stability
  and quasi-local horizons}
\label{sec:definitions}

\subsection{Event horizons}
\label{subsec:eh}

The surface of a black hole is traditionally defined in terms of an
event horizon which is the boundary of the region from where massive
or mass-less particles can reach the outside world.  The formal
definition is however more involved, with the main difficulty being in
how the ``outside world'' is to be defined. It is worthwhile to
briefly sketch the various technical ingredients that go into the
precise formal definition, if only to highlight once again the truly
global nature of event horizons; see
e.g. \cite{Wald:1984,HawkingEllis:1973} for details.

This requires one to attach future and past null-infinity
$\mathscr{I}^\pm$ as boundaries to the physical spacetime and to
consider the causal past of $\mathscr{I}^+$. This causal past is the
region of spacetime $\mathcal{R}$ from which causal signals can escape
to infinity and represents the ``outside-world''.  The future boundary
of $\mathcal{R}$ is the event horizon.  In
Fig.~\ref{fig:schwarzschild}, the region $\mathcal{R}$ is the union of
regions I, II and III, the black hole is of course region II
and the portion of the $r=2M$ surface which divides region II from
I and III is the event horizon.  A little thought shows that in
order for this notion to capture the physical idea we have in mind, it
is necessary to ensure that $\mathscr{I}^+$ is complete in an
appropriate sense.  For example, if we were to look at the causal past
of just a portion of $\mathscr{I}^+$ even for Minkowski space, we
would erroneously conclude the presence of a black hole in Minkowski
space \cite{Geroch:1978ub}. Similarly for Schwarzschild, we could end
up with the wrong location of the event horizon if we looked at only a
portion of $\mathscr{I}^+$ (see again Fig.~\ref{fig:schwarzschild}).

Since we need to construct a complete $\mathscr{I}^+$ and its global
past in this definition, it is clear that the event horizon is a very
global and teleological notion.  There may in fact be an event horizon
forming and growing in the room you are reading this article right now, because
of possible events which might occur a billion years from now. An
example is an observer in the flat region of the Vaidya spacetime in
Fig.~\ref{fig:vaidya}.  This discussion also illustrates that
formally, the notion of an event horizon cannot be used in a
cosmological spacetime such as the one we inhabit, since it fails to
be asymptotically flat.

In practice, the principle of calculating event horizons in numerical
simulations is to start from an educated guess at late times, and to
integrate a null geodesic or a null surface backwards in time
\cite{Thornburg:2006zb,Diener:2003jc}. The basis for these methods is
the fact that when we integrate forwards in time, a small initial
error will cause the null geodesic or surface to diverge exponentially
from the true solution and end up either in the singularity or at
infinity.  This implies that by integrating backwards from even a
reasonable guess at late times one will converge to the true solution
exponentially.

\subsection{Trapped surfaces}
\label{subsec:trapped}

Let's begin with the standard definitions of the $1^{\rm st}$ and
$2^{\rm nd}$ fundamental form of a smooth non-degenerate sub-manifold
$\Ms$ embedded in a spacetime $(\M,g_{ab})$; our discussion mostly
follows \cite{Lee:1997}. The $1^{\rm st}$ fundamental form is just the
induced metric metric $h_{ab}$ on $\Ms$ so that for any vectors $X^a$
and $Y^b$ tangent to $\Ms$:
\begin{equation}
  h_{ab}X^aY^b := g_{ab}X^aY^b\,.
\end{equation}
If $h_{ab}$ is non-degenerate, we can decompose the tangent space
$T_pM$ at any point $p\in\Ms$ as
\begin{equation}
  \label{eq:decomp1}
  T_pM = T_p\Ms\oplus T_p^\perp\Ms\,,\qquad T_p\Ms\cap T_p^\perp\Ms = \{0\}\,,
\end{equation}
where $T_p\Ms$ is the tangent space to $\Ms$ and the subspace
$T^\perp_p\Ms$ is normal to it.  Thus, an arbitrary non-vanishing
vector field $\xi$ defined at points of $\Ms$ can be split uniquely
into a tangential and normal part:
\begin{equation}
  \label{eq:decomp}
  \xi = \xi^\top + \xi^\perp\,\qquad \textrm{where} \qquad \xi^\perp \cdot X = 0
\end{equation}
for any vector $X$ tangent to $\Ms$.  Then, for any $X,Y$ tangent to
$\Ms$ we have
\begin{equation}
  \nabla_XY = (\nabla_XY)^\top  + (\nabla_XY)^\perp\,. 
\end{equation}
The intrinsic covariant derivative on $\Ms$ is then defined as $D_XY
:= (\nabla_XY)^\top$.  The $2^{\rm nd}$ fundamental tensor $\Pi$ is an
operator which takes two vectors tangent to $\Ms$ and produces a
vector in the normal-subspace: 
$T_p\Ms\times T_p\Ms \rightarrow
T^\perp_p\Ms$: 
\begin{equation}
  \Pi(X,Y):=(\nabla_XY)^\perp \,. 
\end{equation}
It is east to show that $D$ defined this way is a legitimate
derivative operator, and that the $2^{\rm nd}$ fundamental form is
symmetric: $\Pi(X,Y) = \Pi(Y,X)$.  The symmetry of $\Pi$ is
especially easy:
\begin{equation}
  \Pi(X,Y) - \Pi(Y,X) = (\nabla_XY - \nabla_YX)^\perp = [X,Y]^\perp = 0\,.
\end{equation}
In the last step we have used the Frobenius theorem which says that
for a smooth sub-manifold $\Ms$, if $X,Y$ are tangent to $\Ms$ then so
is their commutator.

When $\Ms$ is a hyper-surface, i.e. when it has co-dimension 1, then
$N_p$ is 1-dimensional and spanned by the unit-normal $N^a$.  Thus we
can define the \emph{extrinsic curvature} $K_{ab}$ via
\begin{equation}
  \Pi(X,Y)^c:= -(K_{ab}X^aY^b) N^c\,.
\end{equation}
The symmetry of $\Pi$ implies that $K_{ab} = K_{ba}$.  The most
important case for us is however when the sub-manifold $\Ss$ is
2-dimensional and spacelike; the $1^{\rm st}$ fundamental form,
denoted by $q_{ab}$ here, is a Riemannian metric on $\Ss$.  We can
again make the decomposition $T_p\M = T_p\Ss \oplus T_p^\perp\Ss$.
The normal space $T_p^\perp\Ss$ is a $1+1$ dimensional Minkowski
space.  It will be convenient to choose two null vectors $\ell$ and
$n$ to span $T_p^\perp\Ss$. We are of course free to re-scale $\ell$
and $n$ independently by scalars, but we choose to use a normalization
$\ell\cdot n = -1$ which cuts down the re-scaling freedom to
\begin{equation}
  \label{eq:rescaling-ell-n}
 \ell \rightarrow A\ell\,,\quad n \rightarrow A^{-1}n \,.
\end{equation}
We will always choose $(\ell,n)$ to be future directed and $\ell$ and
$n$ as outward and inward pointing respectively. The \emph{mean
  curvature} vector $K^a$ is the trace of the $2^{\rm nd}$ fundamental
form:
\begin{equation}
  K^c := q^{ab}{\Pi_{ab}}^c = -\Theta_{(n)}\ell^c - \Theta_{(\ell)}n^c\,.
\end{equation}
The coefficients appearing here turn out to be precisely the
expansions $\Theta_{(\ell,n)}$ discussed earlier.  Under a re-scaling
of the kind in Eq.~(\ref{eq:rescaling-ell-n}), $K^a$ remains
invariant.  

The different kinds of trapped surfaces correspond to properties of
$K^a$.  Two useful definitions are:
\begin{itemize}
\item \emph{Future trapped surface} ($\Theta_{(\ell)} < 0$,
  $\Theta_{(n)} < 0$): $K^a$ is timelike and future-directed.
\item \emph{Marginally future trapped surface} ($\Theta_{(\ell)} = 0$,
  $\Theta_{(n)} < 0$): $K^a$ is null and future-directed.
\end{itemize}
Furthermore, we shall consider only \emph{closed} surfaces; this is an
important condition as it, among other things, excludes trivial planar
surfaces in flat space.  

It is also useful to remark on the physical significance of the
condition $\Theta_{(n)} < 0$.  This condition holds for round spheres
in flat space and also, as we have seen, for round spheres in the
trapped region of Schwarzschild (see Eq.~(\ref{eq:expansions}).
However, this may not be true for non-spherically symmetric trapped
surfaces even in Schwarzschild.  Furthermore, there are explicit black
hole solutions found by Geroch and Hartle \cite{Geroch:1982bv} which
represent black holes distorted by surrounding matter fields which do
not satisfy $\Theta_{(n)}<0$ at all points on the horizon (but its
average value over the horizon is still negative).  In fact, for a
number of key results that we shall mention below, it is not necessary
to impose $\Theta_{(n)} < 0$. Surfaces with vanishing outward
expansion $\Theta_{(\ell)}=0$, with no restrictions on the sign of
$\Theta_{(n)}$ will be called \emph{marginally outer trapped
  surfaces}, usually abbreviated to MOTS.  A surface with
$\Theta_{(\ell)}<0$ and no condition on $\Theta_{(n)}$ will be said to
be outer trapped. We have seen in the Vaidya example that there are
differences in the location of trapped and outer trapped surfaces.


The collection of closed future-trapped surfaces form the trapped
region of the spacetime. More precisely, the trapped region
$\mathscr{T}$ consists of spacetime points which lie on a closed
future-trapped surface.  The boundary $\mathscr{B}$ of $\mathscr{T}$
is the trapping boundary.  It is also common to consider these concepts
restricted to a spacelike surface $\Sigma$, typically a Cauchy surface
in an initial value problem set-up.  The trapped $\mathscr{T}^\Sigma$
region on $\Sigma$ is the set of points on $\Sigma$ which lie on a
closed future-trapped surface contained entirely on $\Sigma$.  The
boundary of $\mathscr{T}^\Sigma$ is denoted $\mathscr{B}^\Sigma$, and
each connected component of $\mathscr{B}^\Sigma$ is called an
\emph{apparent horizon} if it is the outermost such boundary on
$\Sigma$.  Since $\mathscr{B}^\Sigma$ excludes trapped surfaces not
contained in $\Sigma$, it is clear that $\mathscr{B}^\Sigma \subset
\mathscr{B}\bigcap \Sigma$.  It is important to not confuse the
trapped region $\mathscr{T}$ or its boundary $\mathscr{B}$ with the
black hole region $\mathcal{B}$ defined in Sec.~\ref{subsec:eh} and
the event horizon.  It can be shown \cite{HaywardKriele} that with
some additional regularity conditions, each connected component of the
apparent horizon $\mathscr{B}^\Sigma$ is actually a closed marginally
trapped surface.  The same result was proved in
\cite{Andersson:2007gy} with the regularity assumptions removed.  In
practice, this fact is what is used to locate the apparent horizon in
numerical simulations.

\subsection{The stability of marginally trapped surfaces, trapping and
  dynamical horizons}
\label{subsec:stability}

Let us now go beyond individual trapped/marginally trapped surfaces
and look at their time evolutions.  Consider a region of spacetime
foliated by smooth spacelike surfaces $\Sigma^t$ depending on a real
time parameter $t$.  Start with initial data (the $1^{\rm st}$ and
$2^{\rm nd}$ fundamental forms) at $t=0$ and evolve it using the
Einstein and matter field equations.  This way, we obtain a solution
to the field equations locally in time near $\Sigma^{0}$.  The first
question we wish to address is: If $\Sigma^0$ contains a MOTS $S_0$,
does it persist under time evolution and does it evolve smoothly?  If
it does evolve smoothly, then the union of all the MOTS $S^t$ will
form a smooth 3-surface $\mathcal{H}$ which we shall call a
\emph{marginally trapped tube} (MTT).  A related question is then: How
does $\MTT$ depend on the foliation $\Sigma^t$?  If we start with the
same $\Sigma^0$ but choose the surfaces differently for $t>0$ (still
requiring $\Sigma^t$ to form a smooth foliation), then will we still
end up with a smooth MTT $\MTT^\prime$?  If it exists, is it different
from $\MTT$?  In numerical simulations, it is found that the apparent
horizon can evolve discontinuously.  Can this be understood
analytically?  
\begin{figure}
  \centering
  \includegraphics[width=0.8\textwidth]{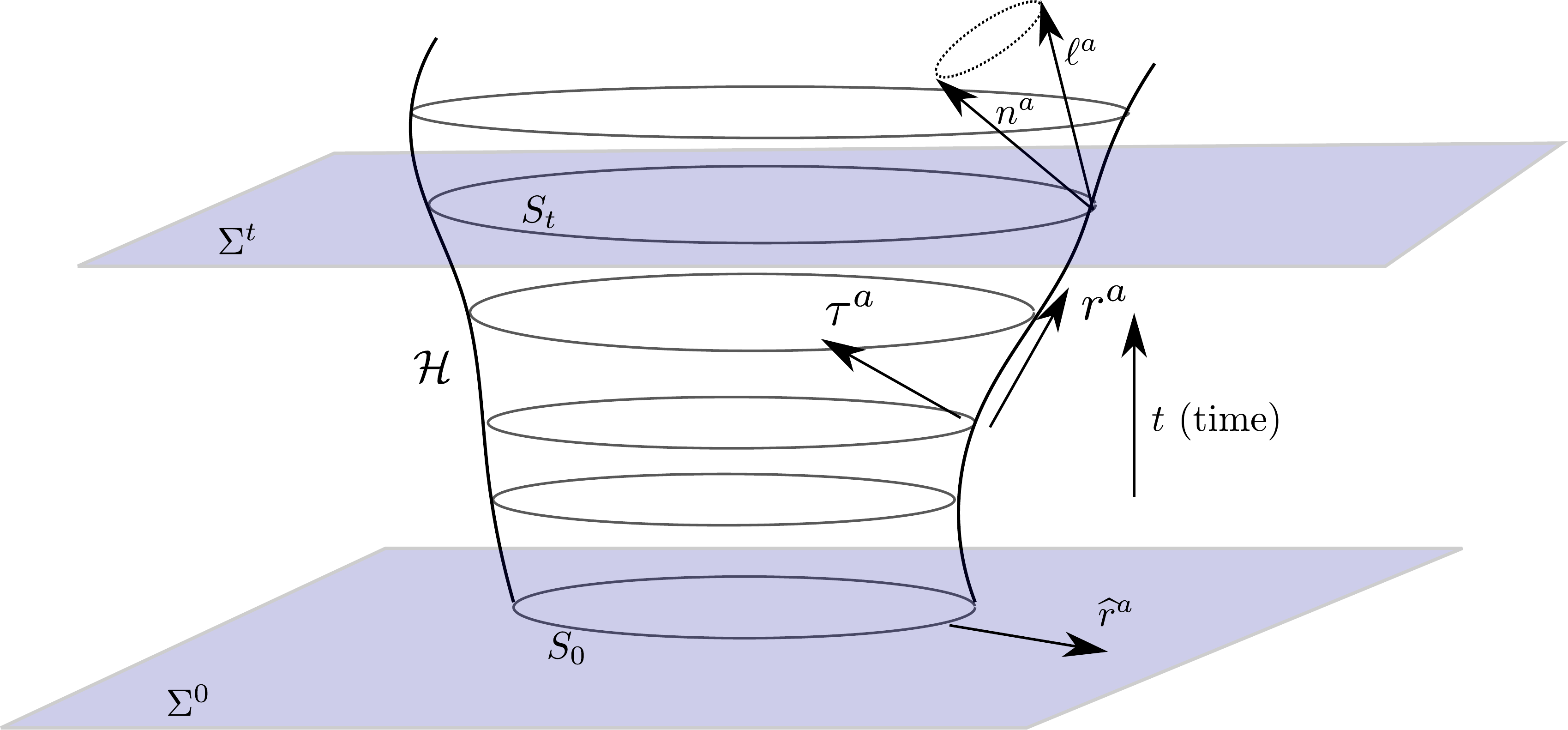}
  \caption{The evolution of a MOTS in time.  Start at $t=0$ with a
    MOTS $S_0$ on a spatial hyper-surface $\Sigma^0$.  Evolve the data on
    $\Sigma^0$ using the Einstein equations.  If the evolution of the
    MOTS is smooth in time, then $S$ will evolve to a MOTS $S_t$ on
    $\Sigma^t$ at time $t>0$.  The collection of all the $S_t$ will
    then form a smooth 3-surface $\mathcal{H}$. We shall see that
    $\mathcal{H}$ will usually be spacelike so that the future null
    vectors $\ell^a$ and $n^a$ orthogonal to $S_t$ (and the future
    null cone) will point ``inwards'' on $\mathcal{H}$.  The vector
    $\widehat{r}^a$ is the unit outwards normal to $S_t$ on
    $\Sigma^t$.}
  \label{fig:mtt}
\end{figure}

\begin{figure}
  \centering
  \includegraphics[width=0.5\textwidth]{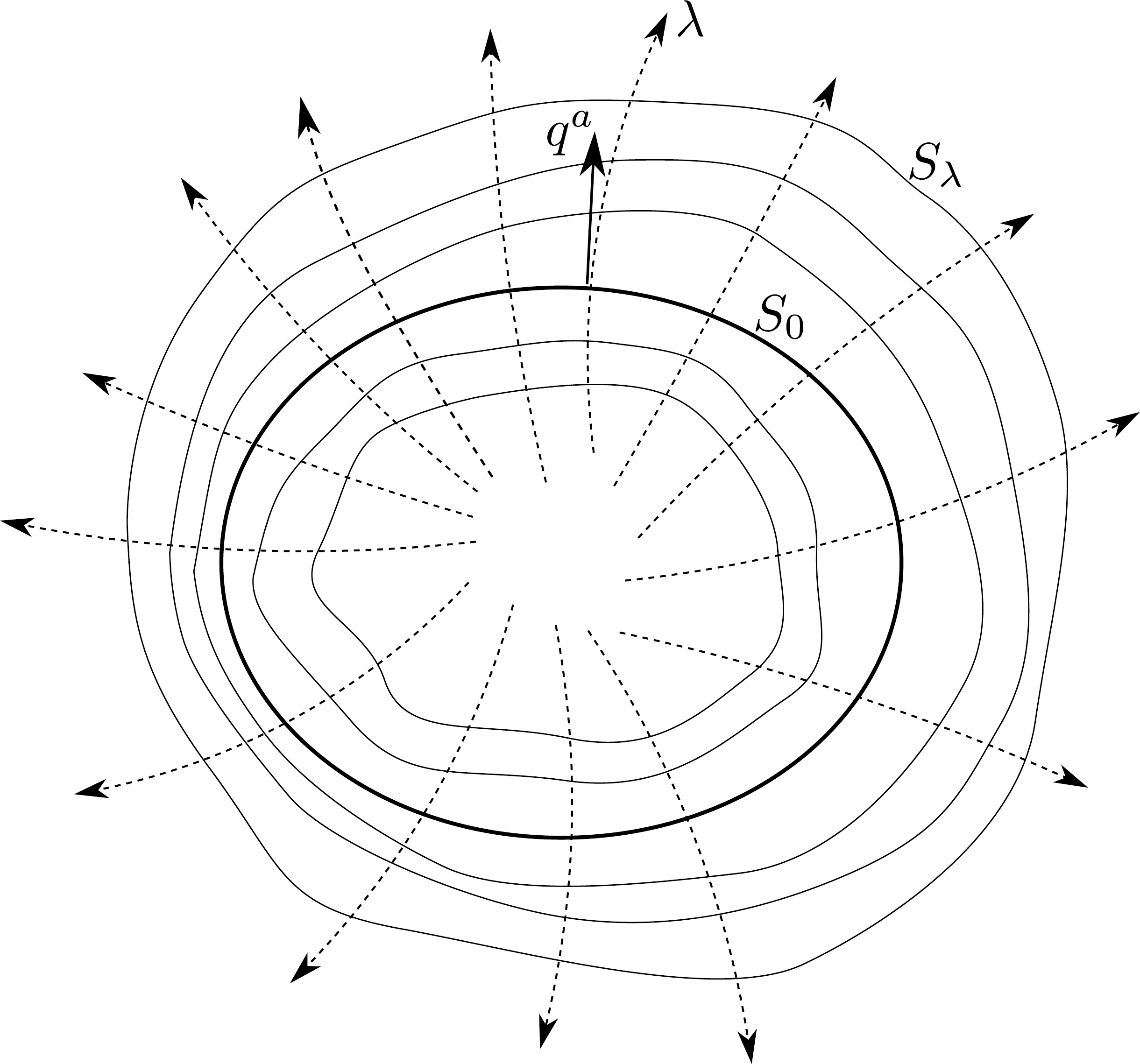}
  \caption{The variation of a surface $S$ viewed as a set of surfaces
    $S_\lambda$.  The initial surface is $S_0$. Each point on $S$
    moves along a smooth curve and $q^a$ is the tangent to these
    curves.}
  \label{fig:variation}
\end{figure}

It turns out that the answer to the above questions are intimately
connected with the \emph{stability} of $S_0$ with respect to
variations on $\Sigma^0$.  To this end, we need to define the notion
of the \emph{geometric variation} of a 2-surface $S$ which is embedded
in a spacetime $\M$
\cite{Newman:1987,Andersson:2008up,Andersson:2007fh,Andersson:2005gq}.
Such a variation is a very general concept; it includes evolving in
time, and an evolution following Einstein equations is a particular
case.  A smooth variation of a sub-manifold $S$ is defined as a
1-parameter family of surfaces $S_\lambda$ (where $\lambda$ is a real
parameter and takes values in some interval $(-\epsilon,\epsilon)$)
such that: $S_0$ is identical to $S$, each $S_\lambda$ is a smooth
surface, and each point on $S$ moves on a smooth curve as $\lambda$ is
varied.  We can then define a vector field $q^a$ as the tangent to
these curves.  This is depicted in Fig.~\ref{fig:variation}.  We could
also perform the variation along null directions or spatial directions
in a given spacetime and of course, the variations do not need to be
uniform on $S$ and different points on $S$ can move at very different
speeds depending on $q^a$.

If we have a relevant geometric quantity on $S$, we can compute it on
each $S_\epsilon$ and differentiate it with respect to $\lambda$, and
the derivative is called the variation of that geometric quantity.  If
$\mathscr{O}_\lambda$ is such a geometric quantity (e.g. the expansion
$\Theta_{(\ell)}$ on each $S_\lambda$), then we shall define
$\delta_q\mathscr{O} := \partial_\lambda
\mathscr{O}_\lambda|_{\lambda=0}$.  It is also important to keep in
mind that for a function $f$, in general the variation is not linear:
$\delta_{fq}\mathscr{O} \neq f\delta_q\mathscr{O}$.

For an MOTS defined on a spatial hyper-surface $\Sigma$, the relevant
variation is along $\widehat{r}^a$, the unit normal to $S$.  The
stability of $S$ is meant to capture the idea that that if $S$ is
deformed outwards, it becomes untrapped, and an inward deformation
leads to $\Theta_{(\ell)}< 0$.  No condition on $\Theta_{(n)}$ is
assumed.  More precisely, the MOTS $S$ is said to be \emph{stably
  outermost} if there exists a function $f\geq 0$ on $S$ such that
$\delta_{f\widehat{r}}\Theta_{(\ell)} \geq 0$.  $S$ is \emph{strictly
  stably outermost} if in addition
$\delta_{f\widehat{r}}\Theta_{(\ell)}\neq 0$ somewhere on $S$.

With this background, we can state the following result
\cite{Andersson:2005gq,Andersson:2007fh}: If $S_0$ is a smooth MOTS on
$\Sigma^0$, and $S_0$ is strictly stably outermost on $\Sigma^0$, then
$S_0$ evolves smoothly into smooth MOTSs $S_t$ on $\Sigma^t$ at time
$t$ at least for sufficiently small (but non-zero) $t$.  Furthermore,
the union of the $S_t$ forms a smooth 3-surface which we shall call
$\mathcal{H}$.  This holds at least as long as the $S_t$ continue to
remain strictly stable outermost.  In addition, if
$G_{ab}\ell^a\ell^b>0$ somewhere on $S$ or if $\ell^a$ has
non-vanishing shear somewhere on $S$, then $\mathcal{H}$ is spacelike.
More generally, if the null energy condition holds then $\mathcal{H}$
is either spacelike or null.

These results answer several of the questions raised at the beginning
of this sub-section.  If we were to start with the same $\Sigma_0$ but
choose different ones at later times, the MTT would still exist at
least for some small time interval since the stability condition still
holds at $t=0$. However, $\mathcal{H}^\prime$ and $\mathcal{H}$ would
not necessarily coincide.  The jumps that are observed in numerical
simulations are because of the outermost condition: while an
individual MTT can continue to evolve smoothly, a new MTT can appear
further outwards.  While not all MOTSs satisfy the stability
condition, in all the cases that the author is aware of in numerical
simulations, the MTTs continue to evolve smoothly. This suggests that
the mathematical results could be strengthened.

Having seen that there is a physically interesting class of MOTSs
which evolve smoothly in time, it is reasonable to impose additional
conditions on an MTT to capture the fact that they are black hole
horizons.  The first is the notion of a \emph{trapping horizon}
defined by Hayward \cite{Hayward:1993wb,Hayward:1994yy} in 1994: A
\emph{future-outer-trapping horizon} (FOTH) is a smooth 3-dimensional
manifold $\mathcal{H}$ foliated by compact 2-manifolds $S_t$
such that
\begin{enumerate}
\item The surfaces $S$ are marginally trapped surfaces
  ($\Theta_{(\ell)}=0$, $\Theta_{(n)}<0$),
\item The directional derivative of $\Theta_{(\ell)}$ along $n^a$
  vanishes: $\Lie_n\Theta_{(\ell)} < 0$.
\end{enumerate}
Historically, this was the first systematic definition of a
quasi-local horizon and it played a key role in spurring further
developments.  At the time, it was not known whether a MOTS would
evolve smoothly in time and if trapped surfaces would limit smoothly
to the boundary of the trapping region (as we have seen in the Vaidya
example, they in fact do not and this possibility was recognized in
\cite{Hayward:1993wb}).  Using this definition, Hayward showed that
$\mathcal{H}$ would be generically spacelike and is null if and only
if the matter flux and the shear of $\ell^a$ vanish identically (the
result from \cite{Andersson:2005gq,Andersson:2007fh} quoted previously
is stronger).  Hayward also showed that it was possible to assign a
surface gravity to the black hole, and to have versions of the laws of
black hole mechanics applicable to $\mathcal{H}$.  The second
condition seems similar to the stability condition defined above, but
in fact the two may not necessarily agree.  A FOTH requires implicitly
that $\ell^a$ and $n^a$ have been extended smoothly in a neighborhood
of $S$ and does not refer to the variation of $S$.

A \emph{dynamical horizon} \cite{Ashtekar:2002ag,Ashtekar:2003hk} is
defined similarly, but it is designed explicitly for the case when the
horizon is spacelike: A smooth 3-dimensional manifold $\mathcal{H}$ is
said to be a \emph{dynamical horizon} (DH) if it is foliated by
compact 2-dimensional manifolds $S_t$ such that
\begin{enumerate}
\item The surfaces $S_t$ are marginally trapped ($\Theta_{(\ell)}=0$,
  $\Theta_{(n)}<0$).
\item $\mathcal{H}$ is spacelike.
\end{enumerate}
The first condition is the same as for a FOTH, but the second
condition specifies \emph{a-priori} the causal nature of $\mathcal{H}$
without any additional conditions on fields transverse to
$\mathcal{H}$.  Even though the definitions of a FOTH and a DH are
similar, neither implies the other.  We shall return to dynamical and
trapping horizons later in Sec.~\ref{sec:dh}, but before that, we
shall discuss the equilibrium case when the MTT is null in some detail.


\section{The equilibrium case: Isolated horizons}
\label{sec:ih}

Having discussed some general properties of trapped surfaces and their
evolutions, we shall now consider a special case that is nevertheless
of significant interest for various applications, namely when the
black hole is in equilibrium with its surroundings and there is no
matter or radiation falling into it.  Since the work of the late 1990s
and the subsequent years, our understanding of this special case has
matured and and we can now consider it be well understood.  All the
well known globally stationary solutions in 4-dimensions, i.e the
Kerr-Newman black holes, are included in this analysis.  Also included
are the dynamical cases when the black hole is itself in equilibrium
but when the time dependent fields are relatively far away from the
black hole.  An example is a system of binary black holes when the
separation of the black holes is much larger than either of the
masses; to a very good approximation each black hole can be treated as
being in equilibrium with its surroundings locally.  In the same
system, once the two black holes have coalesced, the final black hole
will soon reach equilibrium after it's ringdown phase.  The assumption
of global stationarity obviously does not hold in these cases.  Each
of these cases is well modeled by the framework of isolated horizons
that we shall now describe.  We shall start with a prototypical
example namely, the structure of the horizon of a Kerr black hole.
This shall be followed by a general definitions and a summary of some
key results.  We shall conclude with two applications: the near
horizon geometry and the mechanics of isolated horizons.  It will be
convenient to use the Newman-Penrose formalism in this discussion, and
therefore we start with a short digression to discuss this formalism.

We note that strictly speaking, it is not essential to use this
formalism for many of the results that we will discuss later.
However, it does prove to be very useful in a number of cases.  For
our purposes, we discuss it here because it makes calculations very
explicit and is thus useful for pedagogically; there are no tensor
indices and all geometric quantities are expressed in terms of scalar
functions which have clear geometric and physical interpretations.
Since we wish to start with the Kerr black hole as an example, this
formalism is particularly useful for studying its properties
explicitly.

\subsection{The Newman-Penrose formalism}
\label{subsec:np}

The Newman-Penrose formalism \cite{Newman:1961qr,Newman:1962} is a
tetrad formalism where the tetrad elements are null vectors, which
makes it especially well suited for studying null surfaces.  See
\cite{Penrose:1985jw,Chandrasekhar:1985kt,Stewart:1991} for
pedagogical treatments (note that these references take the spacetime
metric to have a signature of $(+---)$ which is different from ours).
Start with a null tetrad $(\ell,n,m,\bar{m})$ where $\ell$ and $n$ are
real null vectors, $m$ is a complex null vector and $\bar{m}$ its
complex conjugate.  The tetrad is such that $\ell\cdot n = -1$,
$m\cdot \bar{m} = 1$, with all other inner products vanishing.  The
spacetime metric is thus given by
\begin{equation}
  \label{eq:51}
  g_{ab} = -\ell_an_b - n_a\ell_b + m_a\bar{m}_b + \bar{m}_am_b\,.
\end{equation}
Directional derivatives along the basis vectors are denoted as
\begin{equation}
  \label{eq:54}
  D := \ell^a\nabla_a\,,\quad \Delta := n^a\nabla_a\,,\quad \delta :=
  m^a\nabla_a\,,\quad \bar{\delta} := \bar{m}^a\nabla_a\,.
\end{equation}
We shall see that the Newman-Penrose formalism employs almost all the
Greek symbols, and therefore often leads to conflicts in notation. An
example is the symbol $\Delta$ which is used for both the directional
derivative along $n^a$ and for the isolated horizon itself; we will
have soon have other such examples.  Hopefully the notation should be
clear from the context.

The components of the connection are encoded in 12 complex scalars,
the spin coefficients, defined via the directional derivatives of the
tetrad vectors:
\begin{EqSystem*}[eq:spincoeffs]
  D\ell& = (\epsilon + \bar{\epsilon})\ell - \bar{\kappa}m -
  \kappa\bar{m} \,,\\
  Dn& = -(\epsilon + \bar{\epsilon})n + \pi m + \bar{\pi}m \,,\\
  Dm& = \bar{\pi}\ell - \kappa n + (\epsilon - \bar{\epsilon})m \,,\\
  \Delta \ell& = (\gamma + \bar{\gamma})\ell - \bar{\tau}m -
  \tau\bar{m}\,,\label{eq:Deltal}\\
  \Delta n & = -(\gamma + \bar{\gamma})n + \nu m + \bar{\nu}\bar{m}\,,\label{eq:Deltan}\\
  \Delta m & = \bar{\nu}\ell - \tau n + (\gamma-\bar{\gamma})m\,,\label{eq:Deltam}\\
  \delta\ell & = (\bar{\alpha} + \beta)\ell -\bar{\rho}m -
  \sigma\bar{m}\,,\\
  \delta n & = -(\bar{\alpha} + \beta)n + \mu m +
  \bar{\lambda}\bar{m}\,,\\
  \delta m & = \bar{\lambda}\ell - \sigma n + (\beta-\bar{\alpha})m\,,\label{eq:deltam}\\
  \bar{\delta}m & = \bar{\mu}\ell - \rho n + (\alpha-\bar{\beta})m\,. \label{eq:deltabarm}
\end{EqSystem*}
While this kind of expansion may seem to some like a backwards step to
the days before efficient tensor notation was developed, it is
actually very convenient in cases where null vectors and null surfaces
are involved.  The use of complex functions is also efficient because
it cuts down the number of quantities by half.  

Many of the spin coefficients have a clear geometric meaning.  First
note that Eq.~(\ref{eq:spincoeffs:a}) implies that $\ell^a$ is
geodesic if and only if $\kappa=0$.  Furthermore, it will be affinely
parameterized if $\epsilon+\bar{\epsilon} = 0$.  Similarly, from
Eq.~(\ref{eq:spincoeffs:e}), we see that $n^a$ is geodesic if and only
if $\nu=0$ and it is affinely parameterized if
$\gamma+\bar{\gamma}=0$.  Important for us in particular are the
coefficients $\rho$ and $\sigma$ which are related to the expansion,
shear and twist defined earlier in Sec.~\ref{subsec:schwarzschild}.
Let $\ell^a$ be tangent to a null geodesic congruence, let it be
affinely parameterized, and let $\zeta^a$ be a connecting vector for
the congruence (it is transverse to $\ell^a$ and satisfies
$[\ell,\zeta]=0$).  As in Eq.~\ref{eq:zetadot}, we need to look at
$\nabla_a\ell_b$ projected in the $(m,\bar{m})$ plane.  Note that in
this plane, the metric is $q^{ab} = 2m^{(a}\bar{m}^{b)}$, the
antisymmetric area form is ${}^2\epsilon = 2im^{[a}m^{b]}$, and the
2-dimensional space of symmetric trace-free $2^{\rm nd}$-rank tensors
is spanned by $m^am^b$ and $\bar{m}^a\bar{m}^b$.  From the definitions
above, it is easy to show that:
\begin{eqnarray}
  \Theta_{(\ell)} &=& q^{ab}\nabla_a\ell_b = m^a\bar{\delta}\ell_a 
  + \bar{m}^a\delta \ell_a = -2\textrm{Re}\rho\,,\\
  m^{[a}\bar{m}^{b]}\nabla_a\ell_b &=& \textrm{Im}\rho\,,\qquad  
  m^am^b\nabla_a\ell_b = -\sigma\,.
\end{eqnarray}
Thus, we see that $-2\textrm{Re}\rho$ is the expansion,
$\textrm{Im}\rho$ is related to the twist, and $\sigma$ is the shear
of $\ell$.  Similarly, real and imaginary parts of $\mu$ gives the
expansion and twist of $n^a$, while $\lambda$ yields its shear.
It is also easy to verify that 
\begin{equation}
  {[}m,\bar{m} {]}^a = (\bar{\mu}-\mu)\ell^a + (\bar{\rho}-\rho)n^a + (\alpha-\bar{\beta})m^a + (\beta-\bar{\alpha})\bar{m}^a\,.
\end{equation}
Thus, using the Frobenius theorem, we see that $m$ and $\bar{m}$ can
be integrated to yield a smooth surface if $\ell^a$ and $n^a$ are
twist free.  Furthermore, since the projection of $\delta m$ is
determined by $\beta-\bar{\alpha}$, it is clear that this determines
the connection, and thus the curvature of this surface.

Since the null tetrad is typically not a coordinate basis, the above
definitions of the spin coefficients lead to non-trivial commutation
relations:
\begin{EqSystem*}[eq:commutation]
  (\Delta D - D\Delta)f &= (\epsilon+\bar{\epsilon})\Delta f + (\gamma
  + \bar\gamma)Df - (\bar\tau + \pi)\delta f -(\tau + \bar{\pi})\bar{\delta}f\,,\\
  (\delta D-D\delta)f &= (\bar{\alpha}+\beta-\bar\pi)Df + \kappa\Delta
  f - (\bar{\rho}+\epsilon-\bar{\epsilon})\delta f
  -\sigma\bar{\delta}f\,,\\
  (\delta\Delta-\Delta\delta)f &= -\bar\nu Df +
  (\tau - \bar{\alpha}-\beta)\Delta f + (\mu-\gamma+\bar\gamma)\delta f + \bar{\lambda}\bar{\delta}f\,,\\
  (\bar{\delta}\delta-\delta\bar{\delta})f &= (\bar{\mu}-\mu)Df +
  (\bar{\rho}-\rho)\Delta f + (\alpha-\bar{\beta})\delta f -
  (\bar{\alpha}-\beta)\bar{\delta}f\,.
\end{EqSystem*}
The Weyl tensor $C_{abcd}$ breaks down into 5 complex scalars
\begin{EqSystem*}
  \Psi_0 = C_{abcd}\ell^am^b\ell^cm^d\,,\quad
  \Psi_1 = C_{abcd}\ell^am^b\ell^cn^d\,,\quad
  \Psi_2 = C_{abcd}\ell^am^b\bar{m}^cn^d\,,\\
  \Psi_3 = C_{abcd}\ell^an^b\bar{m}^cn^d\,,\quad
  \Psi_4 = C_{abcd}\bar{m}^an^b\bar{m}^cn^d\,.
\end{EqSystem*}
Similarly, the Ricci tensor is decomposed into 4 real and 3 complex
scalars $\Phi_{ij}$: 
\begin{EqSystem*}[eq:riccicomp]
  \Phi_{00} = \frac{1}{2}R_{ab}\ell^a\ell^b \,,\quad
  \Phi_{11} = \frac{1}{4}R_{ab}(\ell^a n^b + m^a \bar{m}^b) \,,\quad
  \Phi_{22} = \frac{1}{2}R_{ab}n^a n^b \,,\quad
  \Lambda = \frac{R}{24}\,,\\
  \Phi_{01} = \frac{1}{2}R_{ab}\ell^a m^b \,,\quad
  \Phi_{02} = \frac{1}{2}R_{ab}m^a m^b \,,\quad
  \Phi_{12}  = \frac{1}{2}R_{ab}m^a n^b \,,\quad
  \bar{\Phi}_{ij} = \Phi_{ji}\,.
\end{EqSystem*}
We are allowed to make transformations of the null-tetrad while
preserving their inner products, thereby leading to a representation
of the proper Lorentz group.  The allowed transformations are
parameterized by two real parameters $(A,\psi)$ and two complex
numbers $(a,b)$ (a total of six real parameters in all, as expected)
\begin{description}
\item[(i)] Boosts: $\ell\rightarrow A\ell\,,\quad n\rightarrow A^{-1}n\,,\quad m\rightarrow m$
\item[(ii)] Spin rotations in the $m-\bar{m}$ plane: $
  m\rightarrow e^{i\psi}m\,,\quad \ell\rightarrow \ell\,,\quad
  n\rightarrow n$
\item[(iii)] Null rotations around $\ell$:  $\ell\rightarrow \ell\,,\quad m\rightarrow m + a\ell\,,\quad n
  \rightarrow n + \bar{a}m + a\bar{m} + |a|^2\ell$
\item[(iv)] Null rotations around $n$: $n\rightarrow n\,,\quad m\rightarrow m + b\ell\,,\quad \ell
  \rightarrow \ell + \bar{b}m + b\bar{m} + |b|^2n$
\end{description}
The transformations of the spin coefficients and curvature components
under these transformations are not difficult to work out.  Again, we
refer to \cite{Penrose:1985jw,Chandrasekhar:1985kt,Stewart:1991} for a
more complete discussion.

The relation between the spin coefficients and the curvature
components lead to the so called Newman-Penrose field equations which
are a set of 16 complex first order differential equations.  The
Bianchi identities, $\nabla_{[a}R_{bc]de} = 0$, are written explicitly
as 8 complex equations involving both the Weyl and Ricci tensor
components, and 3 real equations involving only Ricci tensor
components.  See
\cite{Penrose:1985jw,Chandrasekhar:1985kt,Stewart:1991} for the full
set of field equations and Bianchi identities.

\subsection{The Kerr spacetime in the Newman-Penrose formalism}
\label{subsec:kerrnp}

As the prototypical example for an isolated horizon, we now describe
the structure of the Kerr black hole horizon.  This will also
illustrate the utility of the Newman-Penrose formalism when dealing
with null surfaces. A detailed study of the various intricate
properties of the Kerr spacetime can be found in
\cite{Chandrasekhar:1985kt}. Here we shall be brief and focus on the
essential properties of the horizon.

The Kerr metric with mass $M$ and spin $a$ is usually presented in
textbooks as (this is however not the form that Kerr originally
derived it)
\begin{equation}
  ds^2 = -\left(1-\frac{2Mr}{\rho^2}\right)dt^2  + \frac{\rho^2}{\Delta}dr^2 
  -\frac{4aMr\sin^2\theta}{\rho^2}dt\,d\phi + \rho^2d\theta^2 
  + \frac{\Sigma^2\sin^2\theta}{\rho^2}\,d\phi^2
\end{equation}
where 
\begin{equation}
  \rho^2 = r^2 + a^2\cos^2\theta\,, \qquad \Delta = r^2-2Mr+a^2\,,\qquad \Sigma^2 = (r^2+a^2)\rho^2 + 2a^2Mr\sin^2\theta\,.
\end{equation}
This metric has two Killing vectors: a timelike one $\xi^a =
(\partial_v)^a$, and a spacelike rotational one $\varphi^a =
(\partial_\phi)^a$.  Based on the behavior of the metric at large
distances, one can assign a mass $M_\infty=M$ and angular momentum $J_
\infty = aM$ to the spacetime.  There are multiple ways to justify
this.  Because of the existence of the two Killing vectors, the
clearest definition is through the so-called Komar integrals
\cite{Komar:1959} based on the two Killing vectors (see also
\cite{Wald:1984}).  Moreover, again based on the behavior of the
gravitational field at infinity, one can assign two sets of higher
multipole moments $M_k$ and $J_k$
\cite{Geroch:1970cd,Hansen:1974zz,Beig:1980be} which turn out to be
fully determined by $M$ and $a$ : $M_k + iJ_k = M (ia)^k\,,\qquad k =
2,3,\ldots$

As in the original Schwarzschild metric, there are coordinate
singularities when $\Delta = 0$.  This happens when
\begin{equation}
  r = r_\pm = M \pm \sqrt{M^2-a^2}\,.
\end{equation}
These can be removed by a coordinate transformation
$(t,r,\theta,\phi)\rightarrow (v,r,\theta,\varphi)$:
\begin{equation}
  dv = dt + \frac{r^2+a^2}{\Delta}dr\,,\qquad d\varphi = d\phi - \frac{a}{\Delta}dr\,.
\end{equation}
This yields the metric in $(v,r,\theta,\varphi)$ candidate
\begin{eqnarray}
  ds^2 &=& -\left(1-\frac{2Mr}{\rho^2}\right)dv^2 + 2dv\,dr - 2a\sin^2\theta dr\,d\varphi 
  - \frac{4aMr\sin^2\theta}{\rho^2} dv\,d\varphi \nonumber \\
  &+& \rho^2 d\theta^2 +  \frac{\Sigma^2\sin^2\theta}{\rho^2}\,d\varphi^2\,.
\end{eqnarray}
The horizon is the 3-dimensional surface $r=r_+$ which we shall denote
$\Delta$.  The intrinsic metric $q_{ab}$ on $\Delta$ in
$(v,\theta,\varphi)$ coordinates is obtained by setting $r=r_+$ and
dropping the $dr$ terms in this metric.  Rearranging terms we get
\begin{equation}
  q_{ab} = \frac{a^2\sin^2\theta}{\rho_+^2}\left(\nabla_av -\Omega^{-1}d\varphi\right)\left(\nabla_bv -\Omega^{-1}d\varphi\right) + \rho_+^2\nabla_a\theta \nabla_b\theta\,,
\end{equation}
where $\rho_+^2 := r_+^2 + a^2\cos^2\theta$ and $\Omega = a/2Mr_+ =
a/(r_+^2+a^2)$.  It is easy to verify that this metric has signature
$(0++)$ with the degenerate direction being
\begin{equation}
  \label{eq:nullkerr}
  \ell^a\nabla_a = \frac{\partial}{\partial v} + \Omega\frac{\partial}{\partial\varphi}\,.
\end{equation}
Thus, the null normal to $\Delta$ acquires an angular velocity term
in the presence of spin.  

The cross sections of this manifold, i.e. the surfaces of constant
$v$, are spheres with a Riemannian metric $\widetilde{q}_{ab}$.  The
area of such a sphere is time independent: $a_\Delta =
4\pi(r_+^2+a^2)$.  It is easy to verify that if we choose a different
coordinate $v^\prime$ with the cross-sections still being complete
spacelike spheres, the area of each cross section is still $a_\Delta$.

A suitable choice of the ingoing and outgoing future directed null
vectors are
\begin{equation}
  n^a\nabla_a = -\left(\frac{r^2+a^2}{\rho^2}\right)\frac{\partial}{\partial r}\,,\qquad 
  \ell^a\nabla_a = \frac{\partial}{\partial v} + \frac{a}{r^2+a^2}\frac{\partial}{\partial \varphi} + \frac{\Delta}{2(r^2+a^2)}\frac{\partial}{\partial r}\,.
\end{equation}
On $\Delta$ $\ell^a$ agrees with the null direction given in
Eq.~(\ref{eq:nullkerr}).  The other null vector $n^a$ is clearly null
because the metric does not have a $dr^2$ term, and the scalar factor
in $n^a$ is chosen to ensure $\ell\cdot n = -1$.  The covariant
versions are:
\begin{equation}
  n_a = \frac{r^2+a^2}{\rho^2}(-\nabla_av + a\sin^2\theta\nabla_a\varphi) \,,\qquad \ell_a = -\frac{\Delta}{2(r^2+a^2)}\nabla_av + \frac{\rho^2}{r^2+a^2}\nabla_ar + \frac{\Delta a\sin^2\theta}{2(r^2+a^2)} \nabla_a\varphi\,.
\end{equation}
A suitable choice for $m^a$ is 
\begin{eqnarray}
  m_a &=& -\frac{a\sin\theta}{\sqrt{2}\rhot} \nabla_av + \frac{(r^2+a^2)\sin\theta}{\sqrt{2}\rhot}\nabla_a\varphi + \frac{i}{\sqrt{2}}\bar{\rhot}\nabla_a\theta\,,\nonumber \\
  m^a\nabla_a &=& \frac{a\sin\theta}{\sqrt{2}\rhot}\frac{\partial}{\partial v} + \frac{1}{\sqrt{2}\rhot\sin\theta}\frac{\partial}{\partial \varphi} + \frac{i}{\sqrt{2}\rhot}\frac{\partial}{\partial\theta}\,.
\end{eqnarray}
Here we have defined $\rhot := r+ia\cos\theta$, so that $\rho^2 =
|\rhot|^2$.  It is unfortunate that the notation can be confusing.
For example the $\rho^2$ used in the Kerr metric is not to be confused
with the spin coefficient $\rho$.  This should hopefully not cause
confusion because the spin coefficient $\rho$ will vanish identically,
and unless mentioned otherwise, $\rho^2$ will refer to
$r^2+a^2\sin^2\theta$.

We can now compute the spin coefficients at $\Delta$ and for the
moment we shall restrict our attention to those spin coefficients
which are intrinsic to $\Delta$, i.e. do not require any derivatives
transverse to $\Delta$. These are:
$\epsilon,\kappa,\pi,\alpha,\beta,\rho,\sigma,\mu,\lambda$.  As is
typical in tetrad formalisms, we do not need to compute any
Christoffel symbols in order to compute any of the spin coefficients;
the exterior derivative suffices.  Using the definitions of
Eq.~(\ref{eq:spincoeffs}) we can write the exterior derivatives of
$\ell_a, m_a, n_a$ in terms of the exterior products of the basis
vectors.  The spin coefficients are then combinations of contractions
of the exterior derivatives with the basis vectors.  As an example,
the acceleration of $\ell^a$ and its value at the Kerr horizon is
\begin{equation}
  \epsilon + \bar{\epsilon} = \ell^b n^a \cdot 2\nabla_{[a}\ell_{b]} = \frac{r_+-M}{2Mr_+} = \frac{\sqrt{M^2-a^2}}{2M(M+\sqrt{M^2-a^2})},.
\end{equation}
The other spin coefficients at the horizon turn out to be:
\begin{equation}
  \kappa = \sigma = \rho = \lambda = \nu = 0\,, \pi = \alpha+\bar{\beta} = -\frac{\sqrt{2}ar\sin\theta}{\rho^2\bar{\rhot}}
\end{equation}
Some of these can be understood on general grounds.  First, since
$\ell^a$ is tangent to a smooth surface, it must be hyper-surface
orthogonal which means that we must have $\mathrm{Im}\rho =
0$. Furthermore, if a null vector is hyper-surface orthogonal, it can
be shown to be tangent to a geodesic.  This implies $\kappa=0$.  The
important condition on physical grounds is that the expansion of
$\ell^a$ vanishes: $\mathrm{Re}\rho = 0$. Thus, as we saw for a
Schwarzschild black hole, the cross sections of $\Delta$ are
marginally outer trapped surfaces.  Now let us turn to the Weyl
tensor.  It can be shown that the only non-zero component is
\begin{equation}
  \Psi_2 = -\frac{M}{(r-ia\cos\theta)^3}\,.
\end{equation}
Can we now extract from these results general properties of a null
surface which should behave like a black hole horizon in equilibrium?
Can we assign physical quantities such as surface gravity, mass,
angular momentum and higher multipole moments?  We shall answer these
questions in the next sub-section.  Regarding angular momentum, we
note that the spin coefficient which vanishes for $a=0$ is $\pi$.  In
fact, from the values of the spin coefficient, we can check that for
any $X^a$ tangent to the horizon, $X^a\nabla_a\ell^b =
X^a\omega_a\ell^b$ where
\begin{equation}
  \omega_a := -(\epsilon+\bar{\epsilon})n_a + \pi m_a + \bar{\pi}\bar{m}_a\,.
\end{equation}
We shall see that the angular part of $\omega_a$ yields the angular
momentum of the horizon.

\subsection{A primer on null-hyper-surfaces}
\label{subsec:null}

The fundamental geometric objects in the theory of isolated horizons
are null surfaces.  Let us therefore start with with a discussion of
the geometry of null surfaces in a Lorentzian manifold.  The horizon
of a Kerr black hole is a special kind of null surface, namely an
expansion and shear free null surface.  Such null surfaces are rather
special from a geometrical point of view as we shall now explain.

Consider a smooth sub-manifold $\Ms$ of a spacetime $(\M,g_{ab})$.  As
earlier in Sec.~\ref{subsec:trapped}, we define the $1^{\rm
  st}$-fundamental tensor of $\Ms$, i.e. the induced metric $h_{ab}$
as the restriction of $g_{ab}$ to $\Ms$: $h_{ab}X^aY^b :=
g_{ab}X^aY^b$ for any two-arbitrary vector fields $X^a$ and $Y^b$
tangent to $\Ms$.  The sub-manifold $\Ms$ is said to be null when
$h_{ab}$ is degenerate.  In the non-degenerate case, the ambient
covariant derivative operator $\nabla$ induces a natural derivative
operator $\D$ on $\Ms$.  Furthermore, $\D$ is the unique derivative
operator compatible with $h_{ab}$, i.e. $\D_ah_{bc} =0$.

Can we repeat the steps described in Sec.~(\ref{subsec:trapped}) for
defining the fundamental forms and intrinsic connection on a null
surface?  When $\Ms$ is a null hyper-surface, we shall call $\ell^a$ a
null normal if it is along the degenerate direction of $h_{ab}$, so
that $h_{ab}\ell^a = 0$.  We thus encounter a problem in the very
first step, i.e. in the decomposition of Eq.~(\ref{eq:decomp1}). The
null normal $\ell^a$ is also tangent to $\Ms$ so that $T_p\Ms\cap
T_p^\perp\Ms \neq \{0\}$. The other route to defining $\D$, namely
finding the unique derivative operator compatible with $h_{ab}$
doesn't work either because a degenerate metric does not uniquely
determine a derivative operator.  We can still define an intrinsic
derivative operator $\D$ if we pick a particular subspace of $T_pM$
transverse to $\Ms$.  Following \cite{Bejancu:1996}, we first pick a
spacelike subspace $\Sp$ of $T_p\Ms$; for the Kerr horizon, a natural
choice would be vectors tangent to the spherical cross-sections of
fixed $v$.  There will be two 1-dimensional sub-spaces of null vectors
orthogonal to $\Sp$.  One of them is $N_p\Ms$, the null direction of
$h_{ab}$ and the other will be transverse to $\M$ which we shall call
$N_p^\prime\Ms$, and we can decompose $T_p\M$ as $T_p\Ms\oplus
N_p\Ms\oplus N_p^\prime\Ms$.  Associated to a particular null normal
$\ell^a$ we shall pick a vector $n^a$ in the transverse direction by
requiring that $\ell\cdot n = -1$.  We can then decompose $\xi$ as
\begin{equation}
  \xi = \xi^\top + \xi^\perp\qquad\textrm{where}\qquad 
      \begin{array}{ll}
        \xi^\top := \widetilde{\xi} + \alpha\ell \\
        \xi^\perp:= \beta n
      \end{array}\,.
\end{equation}
Here $\alpha$ and $\beta$ are scalars, $\widetilde{\xi}$ is in $\Sp$
at each $p$.  We can then define the intrinsic derivative operator as
before: $\D_XY = (\nabla_XY)^\top$.  However, $\D$ would depend on our
choice of $\Sp$ and unlike in the non-degenerate case, there is in
general no natural canonical choice.  There is however one case when
this is not an issue, namely when the $2^{\rm nd}$-fundamental form
vanishes, $\nabla_XY$ is always tangential to $\Ms$ and there is no
need to decompose $\nabla_XY$.  This happens when:
\begin{equation}
  \label{eq:xdy0}
  \ell_aX^b \nabla_b Y^a = -X^aY^b\nabla_a\ell_b = 0
\end{equation}
for any $X^a, Y^a$ tangent to $\Ms$.  Thus, $\ell_a$ is covariantly
constant on the null surface. 

An alternative way to state the same result (emphasized in
e.g. \cite{Ashtekar:2001is}) is: if we start with a vector tangent to
$\Delta$ and parallel transport it using the spacetime derivative
operator $\nabla$ along a curve lying on $\Delta$, then the vector
remains tangent to $\Delta$.  Since parallel transport of $Y^a$ along
$X^a$ is defined by $X^a\nabla_aY^b = 0$, this alternative criteria is
also equivalent to $\ell_aX^b\nabla_bY^a = 0$.  We shall see that
Eq.~(\ref{eq:xdy0}) is satisfied for the various kinds of horizons
that we shall now define.

\subsection{Non-expanding, weakly-isolated and isolated horizons}
\label{subsec:ihdefs}

Having understood null-surfaces, we are now ready to define different
kinds of isolated horizons with increasingly stronger conditions.  We
shall start with the minimum set of conditions, namely a marginally
trapped tube which is null, and no condition on the ingoing expansion.
A smooth 3-dimensional null surface $\Delta$ is said to be a
\emph{non-expanding horizon} if:
\begin{itemize}
\item $\Delta$ has topology $S^2\times\mathbb{R}$, and if $\varpi: S^2
  \times \mathbb{R} \rightarrow S^2$ is the natural projection, then
  $\varpi^{-1}(x)$ for any $x\in S^2$ are null curves on $\Delta$.
\item The expansion $\Theta_{(\ell)} := q^{ab}\nabla_a\ell_b$ of any
  null normal $\ell^a$ of $\Delta$ vanishes.
\item The Einstein field equations hold at $\Delta$, and the matter
  stress-energy tensor $T_{ab}$ is such that for any future directed
  null-normal $\ell^a$, $-T^a_b\ell^b$ is future causal.  
\end{itemize}
We shall consider only null tetrads adapted to $\Delta$ such that, at
the horizon, $\ell^a$ coincides with a null-normal to $\Delta$.  We
shall also consider a foliation of the horizon by spacelike spheres
$S_v$ with $v$ a coordinate on the horizon which is also an affine
parameter along $\ell$: $\mathcal{L}_\ell v = 1$; $S$ shall denote a
generic spherical cross-section of $\Delta$. Null rotations about
$\ell^a$ correspond to changing the foliation.

This deceptively simple definition of a non-expanding horizon leads to
a number of important results which we state here without proof, most
of which are however well illustrated by the Kerr example discussed
earlier.
\begin{enumerate}
\item Any null normal $\ell^a$ is a symmetry of the intrinsic
  degenerate metric $q_{ab}$ on $\Delta$: $\Lie_\ell q_{ab} = 0$.  
\item The null normal of $\Delta$ is only given to be expansion free.
  However, a non-expanding horizon is also shear free.  To show this,
  we use the Raychaudhuri equation:
  \begin{equation}
    \Lie_\ell\Theta_\ell = \kappa_{\ell}\Theta_{(\ell)} - \frac{1}{2}\Theta_{(\ell)}^2 - |\sigma|^2 - R_{ab}\ell^a\ell^b\,. 
  \end{equation}
  Setting $\Theta_{(\ell)}=0$, and observing that the energy condition
  implies $R_{ab}\ell^a\ell^b\geq 0$, we get that the sum of two
  non-negative quantities must vanish.
  \begin{equation}
    |\sigma|^2 + R_{ab}\ell^a\ell^b = 0\,.
  \end{equation}
  This can only happen if $\sigma=0$ and $R_{ab}\ell^a\ell^b=0$ on the
  horizon.  Thus, the full projection of $\nabla_a\ell_b$ on $\Delta$
  vanishes, and as we saw in Sec.~\ref{subsec:null}, this is just the
  condition required to ensure that the induced derivative operator on
  $\Delta$ is well defined.
\item The Weyl tensor components $\Psi_0$ and $\Psi_1$ vanish on the
  horizon.  This implies that $\Psi_2$ is an invariant on $\Delta$ as
  long as the null-tetrad is adapted to the horizon; it is
  automatically invariant under boosts and spin rotations (it has spin
  weight $0$), and it is invariant under null rotations around $\ell$
  because $\Psi_0$ and $\Psi_1$ vanish.  Similarly, the Maxwell field
  component $\phi_0$ vanishes on the horizon, and $\phi_1$ is
  invariant on $\Delta$.  Both $\Psi_2$ and $\phi_1$ are also time
  independent on the horizon.    
\item There exists a 1-form $\omega_a$ such that, for any vector field
  $X^a$ tangent to $\Delta$,
  \begin{equation}
    \label{eq:14}
    X^a\nabla_a\ell^b = X^a\omega_a\ell^b\,.
  \end{equation}
  The 1-form $\omega_a$ plays a fundamental role in what follows.  The
  pullback of $\omega_a$ to the cross-sections $S$ will be denoted
  $\tilde{\omega}_a$.
\item The surface gravity of $\ell$ is
  \begin{equation}
    \label{eq:15}
    \tilde{\kappa}_{(\ell)} = \ell^a\omega_a\,.
  \end{equation}
  We will say that $\Delta$ is extremal if $\tilde{\kappa}_{(\ell)}=0$
  and non-extremal otherwise.  Here we shall always assume that
  $\Delta$ is non-extremal.  The curl and divergence of $\omega$ carry
  important physical information.  The curl is related to the
  imaginary part of the Weyl tensor on the horizon
  \begin{equation}
    \label{eq:16}
    d\omega = \textrm{Im}\left[\Psi_2\right] {}^{2}\epsilon\,.
  \end{equation}
  and its divergence specifies the foliation of $\Delta$ by spheres
  \cite{Ashtekar:2001jb}.
\item By the geometry of $\Delta$, we shall mean the pair
  $(q_{ab},\D_a)$.  Clearly, $q_{ab}$ yields a Riemannian metric
  $\widetilde{q}_{ab}$ on the cross-sections of $\Delta$. In turn,
  $\D_a$ is determined by $\omega_a$ and by the unique derivative
  operator $\widetilde{\D}_a$ compatible with $\widetilde{q}_{ab}$.  
\end{enumerate}
We need to strengthen the conditions of a non-expanding horizon for
various physical situations.  The minimum extra condition required for
black hole thermodynamics and to have a well defined action principle
with $\Delta$ as an inner boundary of a portion of spacetime, is
formulated as a weakly isolated horizon \cite{Ashtekar:2000hw}: A
weakly isolated horizon $(\Delta,[\ell])$ is a non-expanding horizon
equipped with an equivalence class of null normals $[\ell]$ related by
constant positive rescalings and such that 
\begin{equation}
  \label{eq:39}
  \mathcal{L}_\ell\omega_a = 0\,.
\end{equation}
If we recsale $\ell\rightarrow f\ell$, $\omega_a$ transforms as
$\omega_a\rightarrow \omega_a + \partial_a\ln f$.  It is thus
invariant under constant rescalings and there is a unique $\omega_a$
corresponding to the equivalence class $[\ell]$.

The zeroth law holds on weakly isolated horizons,
i.e. $\tilde{\kappa}_{(\ell)} = \ell^a\omega_a$ is constant on
$\Delta$:
\begin{equation}
  \Lie_\ell \omega_a = \ell^b 2\D_{[b}\omega_{a]} + \D_a(\ell^b\omega_b) = \D_a\tilde{\kappa}_{(\ell)} \,.
\end{equation}
In the second step we have used Eq.~(\ref{eq:16}) to conclude that
$\ell^b \times 2\D_{[b}\omega_{a]} =
\textrm{Im}\ell^b\left[\Psi_2\right] {}^{2}\epsilon_{ba} = 0$.  Note
that under a re-scaling $\ell^a \rightarrow f\ell^a$, $\omega_a$
transforms as $\omega_a \rightarrow \omega_a + D_a\ln f$ so that it is
invariant under constant rescalings.

Any non-expanding horizon can be made into a weakly isolated horizon
by suitably scaling the null generators.  Thus, the restriction to
weakly isolated horizons is not a genuine physical restriction.  One
could go ahead and impose further physical restrictions on the
intrinsic horizon geometry by requiring that not only $\omega_a$, but
the full connection $\D_a$ on $\Delta$ is preserved by $\ell^a$: An
isolated horizon $(\Delta,[\ell])$ is thus a non-expanding horizon
equipped with a equivalence class of null-normals related by constant
positive rescalings such that $[\mathcal{L}_\ell,\D] = 0$
\cite{Ashtekar:2001jb}.

It can be shown that the gauge-invariant geometry of an axisymmetric
isolated horizon can be fully specified by the area $a_\Delta$ and two
set of multipole moments $M_n,J_n$ for $n=0,1,2,\ldots$
\cite{Ashtekar:2004gp}.  In contrast to the field multipole moments
defined at infinity, $M_n$ and $J_n$ are the source multipole moments.
We shall not discuss these moments here in any detail, except to say
that in the vacuum case, the moments are essentially obtained by
decomposing $\Psi_2$ at $\Delta$ into spherical harmonics based on
preferred coordinates adapted to the axial symmetry. As at infinity,
the Kerr horizon corresponds to a specific choice of these multipoles,
but due to the non-linearity of general relativity, the field and
source moments will not generally agree.  See
e.g. \cite{Schnetter:2006yt,Jasiulek:2009zf,Jaramillo:2011re,Saijo:2011ag}
for some applications of these multipole moments.

\subsection{The near horizon geometry}
\label{subsec:nearhorizon}

In a number of astrophysical applications where black holes play a
role, it is not directly the horizon which is involved but rather the
spacetime in the vicinity of the black hole.  An especially
interesting example of relevance to gravitational wave observations is
the case of a binary system consisting of a stellar mass black hole or
star orbiting around a much larger black hole (see
e.g. \cite{AmaroSeoane:2007aw} for a review of the astrophysics of
such systems).  As the small particle orbits the large black hole, it
effectively ``maps'' the spacetime in the vicinity of the large black
hole, much as the motion of satellites around Earth enables us to map
Earth's gravitational potential and therefore its shape.  If carried
out to sufficient precision, a measurement of gravitational waves from
such a system would enable us to determine the gravitational field in
the vicinity of the large black hole (see
e.g. \cite{Ryan:1997hg,Hughes:2000ssa}).  One expects that the black
hole is well approximated by a Kerr spacetime and we can seek to
measure deviations from it thereby testing an important prediction of
general relativity.  Typical studies on this subject assume the black
hole to be Kerr, which is entirely reasonable to a very good
approximation (nearby stars or other matter fields might distort the
black hole somewhat, but this is expected to be a small
effect). However, for mathematical purposes, we could pose the
question from a different viewpoint: If we specify the intrinsic
geometry of the horizon, then to what extent is the near horizon
spacetime determined by the horizon geometry?  If we assume the large
black hole to be in equilibrium (an excellent approximation for the
case we have just described), then it seems reasonable to model the
black hole as an isolated horizon. So the question then is: Can we
find solutions to the Einstein field equations which admit a generic
isolated horizon as an inner boundary?  If so, then what is the extra
data (beyond the intrinsic horizon geometry) that needs to be
specified?  (A full solution to the problem would require us to go
beyond isolated horizons an to consider small deviations from
equilibrium, but we shall not discuss this generalization here).

It turns out that these questions can be clearly answered if we use
the characteristic initial value formulation of Einstein's equations
where free data is specified on a set of intersecting null
hyper-surfaces
\cite{Friedrich:1981at,Rendall:1990,Friedrich:2000qv,Stewart:1991}.
Consider $N$ dependent variables $\psi_I (I=1,\ldots,N)$ on a
spacetime manifold with coordinates $x^a$. We shall be concerned with
hyperbolic first-order quasi-linear equations of the form
\begin{equation}
  \sum_{J=1}^N A^a_{IJ}(x,\psi)\partial_a\psi_J + F_I(x,\psi) = 0\,.
\end{equation}
In the standard Cauchy problem, one specifies the $\psi_I$ at some
initial time.  A solution is then guaranteed to be unique and to exist
at least locally in time.  The characteristic formulation considers
a pair of null surfaces $\mathcal{N}_0$ and $\mathcal{N}_1$ whose
intersection is a co-dimension-2 spacelike surface $S$.  It turns out
to be possible to specify appropriate data on the null surfaces and on
$S$ such that the above system of equations is well posed and has a
unique solution, at least locally near $S$.

\begin{figure}
  \centering
  \includegraphics[width=0.7\textwidth]{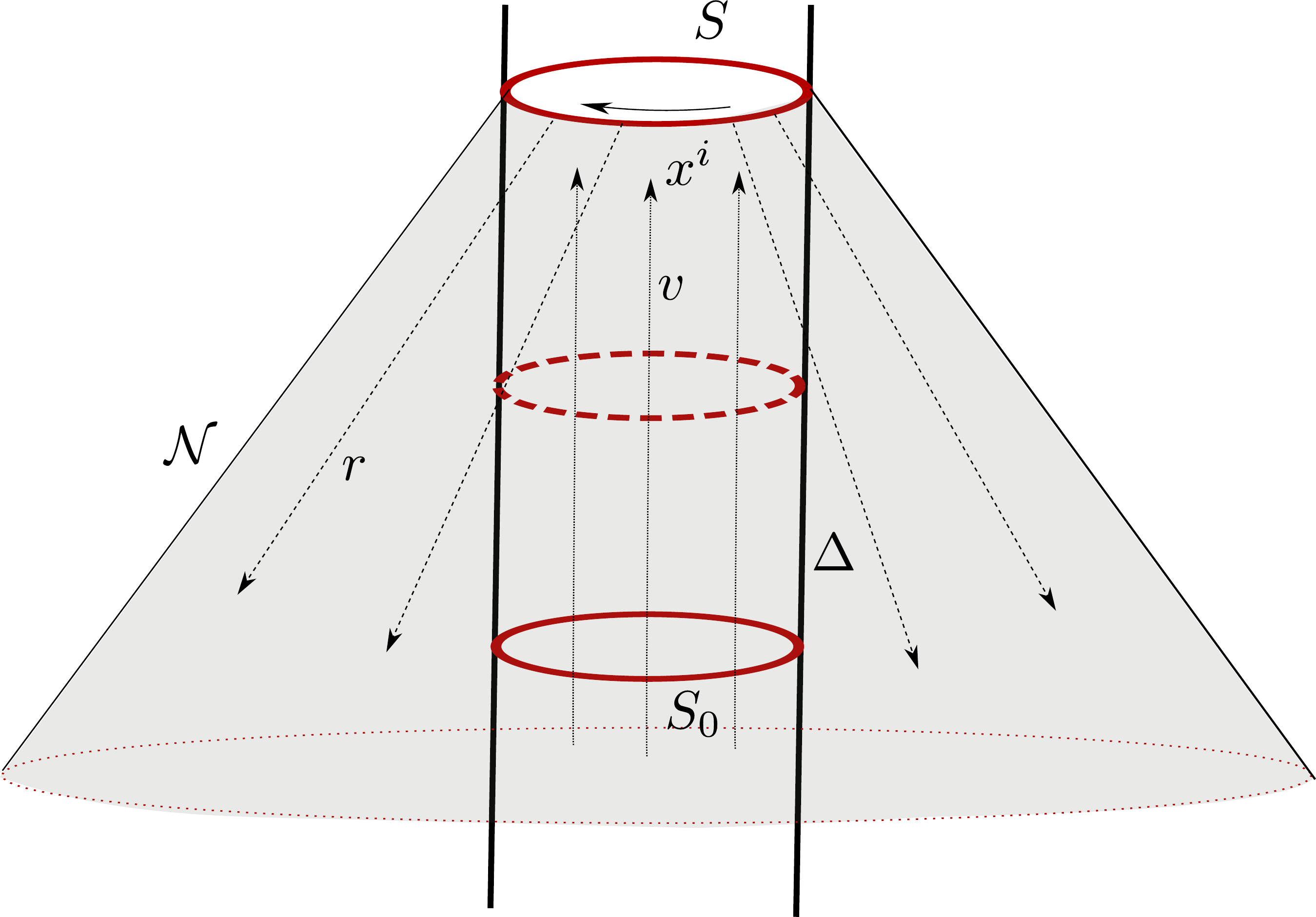}
  \caption{The near horizon coordinates.  The isolated horizon is
    $\Delta$ and the transverse null surface is $\mathcal{N}$. The
    affine parameter along the outgoing null geodesics on
    $\mathcal{N}$ is $r$, and $v$ is a coordinate along the null
    generators on $\Delta$, and $x^1$ are coordinates on the
    cross-sections of $\Delta$.  }
  \label{fig:nearhorizon}
\end{figure}

In our case, the appropriate free data is specified on the horizon and
on an outgoing past light cone originating from a cross section of the
horizon.  Such a construction in the context of isolated horizons was
first studied by Lewandowski \cite{Lewandowski:1999zs} who
characterized the general solution of Einstein equations admitting an
isolated horizon.  This was worked out in detail in
\cite{Krishnan:2012bt} which we follow here; similar results from a
somewhat different perspective are discussed in \cite{Booth:2012xm}.
The general scenario is sketched in Figure~\ref{fig:nearhorizon}.  We
consider a portion of the horizon $\Delta$ which is isolated, in the
sense that no matter and/or radiation is falling into this portion of
the horizon.  For a cross-section $S$, the past-outgoing light cone is
denoted by $\mathcal{N}$. The null generators of $\Delta$ and
$\mathcal{N}$ are parameterized by $v$ and $r$ respectively; $x^i$ are
coordinates on $S$.  This leads to a coordinate system $(v,r,x^i)$
which is valid till the null geodesics on $\mathcal{N}$ start to
cross.  The field equations are solved in a power series in $r$ away
from the horizon.

Let us now assume that the vacuum Einstein equations hold in a
neighborhood of the horizon $\Delta$.  Following
\cite{Ashtekar:2000sz} we introduce a coordinate system and null
tetrad in the vicinity of $\Delta$ analogous to the Bondi coordinates
near null infinity.  See Fig.~\ref{fig:nearhorizon}.  Choose a
particular null normal $\ell^a$ on $\Delta$. Let $v$ be the affine
parameter along $\ell^a$ so that $\ell^a\nabla_a v = 1$.  Let $S_v$
denote the spheres of constant $v$. Introduce coordinates $x^i$
($i=2,3$) on any one $S_v$ (call this sphere $S_0$) and require them
to be constant along $\ell^a$: $\ell^a\nabla_a x^i = 0$; this leads to
a coordinate system $(v,x^i)$ on $\Delta$.  Let $n^a$ be a future
directed inward pointing null vector orthogonal to the $S_v$ and
normalized such that $\ell\cdot n = -1$. Extend $n^a$ off $\Delta$
geodesically, with $r$ being an affine parameter along $-n^a$; set
$r=0$ at $\Delta$.  This yields a family of null surfaces
$\mathcal{N}_v$ parameterized by $v$ and orthogonal to the spheres
$S_v$.  Set $(v,x^i)$ to be constant along the integral curves of
$n^a$ to obtain a coordinate system $(v,r,x^i)$ in a neighborhood of
$\Delta$.  Choose a complex null vector $m^a$ tangent to $S_0$.  Lie
drag $m^a$ along $\ell^a$:
\begin{equation}
  \label{eq:17}
  \mathcal{L}_\ell m^a = 0 \,\quad \textrm{on}\,\Delta\,.
\end{equation}
We thus obtain a null tetrad $(\ell,n,m,\bar{m})$ on
$\Delta$. Finally, parallel transport $\ell$ and $m$ along $-n^a$ to
obtain a null tetrad in the neighborhood of $\Delta$.  This
construction is fixed up to the choice of the $x^i$ and $m^a$ on an
initial cross-section $S_0$.  We are allowed to perform an arbitrary
spin transformation $m\rightarrow e^{i\psi}m$ on $S_0$.

With the Bondi-like coordinate system in hand, we can now in principle
use the coordinate basis vectors in the $(v,r,x^i)$ coordinates to
construct an arbitrary null-tetrad near the horizon.  The evolution
equations for the component functions of the null tetrad will follow
from the above construction.  Let us start with $n_a$ and $n^a$.  We
have the family of null surfaces $\mathcal{N}_v$ parameterized by $v$;
$n_a$ is normal to the $\mathcal{N}_v$, and $r$ is an affine parameter
along $-n^a$.  This implies that we can choose
\begin{equation}
  \label{eq:2}
  n_a = -\partial_a v \qquad \textrm{and} \qquad n^a\nabla_a
  :=\Delta = -\frac{\partial}{\partial r}\,.
\end{equation}
To satisfy the inner-product relations $\ell^a n_a = -1$ and
$m^an_a=0$, the other basis vectors must be of the form:
\begin{equation}
  \label{eq:3}
  \ell^a\nabla_a := D = \frac{\partial}{\partial v} + U\frac{\partial}{\partial r}
  + X^i\frac{\partial}{\partial x^i}\,,\quad m^a\nabla_a:= \delta =
  \Omega\frac{\partial}{\partial r} + \xi^i\frac{\partial}{\partial x^i}\,.
\end{equation}
The frame functions $U,X^i$ are real while $\Omega,\xi^i$ are complex.
We wish to now specialize to the case when $\ell^a$ is a null normal
of $\Delta$ so that the null tetrad is adapted to the horizon.  Since
$\partial_v$ is tangent to the null generators of $\Delta$, this
clearly requires that $U,X^i$ must vanish on the horizon.  Similarly,
we want $m^a$ to be tangent to the spheres $S_v$ at the horizon, so
$\Omega$ should also vanish on $\Delta$.  Thus, $U,X^i,\Omega$ are all
$\mathcal{O}(r)$ functions.

To expand the metric in powers of $r$, we start with the frame fields,
and the radial frame equations derived from the
commutation relations.  The strategy is the same as for the spin
coefficients.  The radial equations give us the first radial
derivatives by substituting the horizon values on the right hand side,
and taking higher derivatives leads to the higher order terms. The
calculations are straightforward and lead to the following expansions:
\begin{EqSystem*}[eq:27]
  U = r\tilde{\kappa} + r^2\left(2\left|\pi^{(0)}\right|^2 +
    \mathrm{Re}[\Psi_2^{(0)}]\right) + \mathcal{O}(r^3)\,, \\
  \Omega = r\bar{\pi}^{(0)} + r^2\left( \mu^{(0)}\bar{\pi}^{(0)} +
    \bar{\lambda}^{(0)}\pi^{(0)} + \frac{1}{2}\bar{\Psi}_3^{(0)}\right) +
  \mathcal{O}(r^3)\,, \\
  X^i = \bar{\Omega} \xi^i_{(0)}  + \Omega\bar{\xi}^i_{(0)} + \mathcal{O}(r^3)\,,\\
  \xi^i = \left[ 1 + r\mu^{(0)} + r^2\left((\mu^{(0)})^2+|\lambda^{(0)}|^2\right)\right]\xi^i_{(0)}
  \nonumber \\ \qquad + \left[ r\bar{\lambda}^{(0)}  + r^2\left( 2\mu^{(0)}\bar{\lambda}^{(0)} +
      \frac{1}{2}\bar{\Psi}_4^{(0)}\right) \right]\bar{\xi}^i_{(0)}  +
  \mathcal{O}(r^3)\,.
\end{EqSystem*}
The contravariant metric is seen to be given in terms of the frame fields
as follows:
\begin{EqSystem*}[eq:28]
  g^{rr} = 2(U+|\Omega|^2) \,,\qquad
  g^{vr} = 1\,,\\
  g^{ri} = X^i + \bar{\Omega}\xi^i + \Omega\bar{\xi}^i \,,\qquad
  g^{ij} = \xi^i\bar{\xi}^j + \bar{\xi}^i\xi^j\,. 
\end{EqSystem*}
The null co-tetrad can be calculated easily up to $\mathcal{O}(r^2)$ :
\begin{EqSystem}
  n = -dv\,,\\
  \ell = dr - \left(\tilde{\kappa}r +
    \mathrm{Re}[\Psi_2^{(0)}]r^2\right)dv - \left(\pi^{(0)} r +
    \frac{1}{2}\Psi_3^{(0)}r^2\right)\xi^{(0)}_idx^i - \left(\bar{\pi}^{(0)} r
    +
    \frac{1}{2}\bar{\Psi}_3^{(0)}r^2\right)\bar{\xi}^{(0)}_idx^i\,,\\
  m = -\left(\bar{\pi}^{(0)}r+ \frac{1}{2}\Psi_3^{(0)}r^2\right)dv +
  (1-\mu^{(0)}r)\xi^{(0)}_idx^i - \left(\bar{\lambda}^{(0)}r +
    \frac{1}{2}\bar{\Psi}_4^{(0)}r^2\right)\bar{\xi}^{(0)}_idx^i\,. 
\end{EqSystem}
Here $\xi^{(0)}_i$ are defined by the relations $\xi^{(0)}_i\xi^i_{(0)} = 0$ and
$\xi^{(0)}_i\bar{\xi}^i_{(0)} = 1$; it will be convenient to set $m^{(0)}_a :=
\xi^{(0)}_i\partial_ax^i$.  In powers of $r$, the metric is:
\begin{equation}
  \label{eq:metric-expansion}
  g_{ab} = -2\ell_{(a}n_{b)} + 2m_{(a}\bar{m}_{b)} = g_{ab}^{(0)} +
  g_{ab}^{(1)}r + \frac{1}{2}g_{ab}^{(2)}r^2 + \cdots\,,
\end{equation}
where
\begin{Equation}
  \label{eq:metric0}
  g_{ab}^{(0)} = 2\partial_{(a}r\partial_{b)}v +
  2m^{(0)}_{(a}\bar{m}^{(0)}_{b)}\,,
\end{Equation}
\begin{MultiLine}
  \label{eq:metric1}
  g_{ab}^{(1)} =  -\left( 2\tilde{\kappa}\partial_{(a}v\partial_{b)}v +
    4\pi^{(0)}m^{(0)}_{(a}\partial_{b)}v +
    4\bar{\pi}^{(0)}\bar{m}^{(0)}_{(a}\partial_{b)}v \right. \\ +
    \left. 4\mu^{(0)}m^{(0)}_{(a}\bar{m}^{(0)}_{b)} +
    2\lambda^{(0)}m^{(0)}_{(a}m^{(0)}_{b)} +
    2\bar{\lambda}^{(0)}\bar{m}^{(0)}_{(a}\bar{m}^{(0)}_{b)}\right)\,,
\end{MultiLine}
\begin{MultiLine}
  \label{eq:metric2}
  g_{ab}^{(2)} = 4\left(|\pi^{(0)}|^2 -
    \mathrm{Re}[\Psi_2^{(0)}]\right)\partial_{(a}v\partial_{b)}v 
  + 4\left( \mu^{(0)}\pi^{(0)} + \lambda^{(0)}\bar{\pi}^{(0)}
    -\Psi_3^{(0)}\right)m^{(0)}_{(a}\partial_{b)}v \\*
  +4\left( \mu^{(0)}\bar{\pi}^{(0)} + \bar{\lambda}^{(0)}\pi^{(0)} -
    \bar{\Psi}_3^{(0)}\right)\bar{m}^{(0)}_{(a}\partial^{}_{b)}v 
  + 4\left((\mu^{(0)})^2+\left|\lambda^{(0)}\right|^2 \right)
  m^{(0)}_{(a}\bar{m}^{(0)}_{b)} \\*
  + \left(4\mu^{(0)}\lambda^{(0)} - 2\Psi_4^{(0)}
  \right)m^{(0)}_{(a}m^{(0)}_{b)} +
  \left(4\mu^{(0)}\bar{\lambda}^{(0)} -2\bar{\Psi}_4^{(0)}
  \right)\bar{m}^{(0)}_{(a}\bar{m}^{(0)}_{b)}\,.
\end{MultiLine}
Iterating this procedure to higher orders is, in principle,
straightforward.  This calculation provides a starting point for a
number of ongoing work in applying isolated and quasi-local horizons
to astrophysical situations.

\subsection{Angular momentum, mass, and the first law for isolated
  horizons}
\label{subsec:firstlawih}

The first law for black holes, and black hole thermodynamics in
general, was developed by Bekenstein \cite{Bekenstein:1973ur}, and by
Bardeen, Carter and Hawking \cite{Bardeen:1973gs} in 1973 in analogy
with the laws of thermodynamics.  The he zeroth law says that the
surface gravity is constant over the black hole horizon.  We have
already seen that this is true for a weakly isolated horizon: the
surface gravity $\kappa_{\ell} = \ell^a\omega_a$ is constant on
$\Delta$.  The main difference with the standard formulation for a
globally stationary the surface gravity refers to the globally defined
stationary Killing vector which is normalized to have unit norm at
infinity.  A weakly isolated horizon does not refer to any globally
defined Killing vector.  The standard formulation of the second law
says that the area of the event horizon can never decrease in time.
The area of a non-expanding horizon is constant, thus the second law is
trivial in this context.

Let us now turn to the first law.  The standard formulation for a
stationary black hole is
\begin{equation}
  \delta M = \frac{\kappa}{8\pi}\delta a + \Omega\delta J\,.
\end{equation}
Here $M$ is the mass measured at spatial infinity, $\kappa$ is the
surface gravity at the event horizon but using a vector field
normalized at infinity, $a$ is the area of the horizon, $\Omega$ is
the angular velocity at the horizon, $J$ is the angular momentum at
infinity.  Electromagnetic fields can also be included and leads to
additional terms This is reasonable when applied to a globally
stationary spacetime, but clearly needs to be refined for a black hole
in equilibrium locally in an otherwise dynamical spacetime.  It turns
out that it is possible to formulate the first law for isolated
horizons using quantities defined only at the horizon, without
reference to the behavior of fields at infinity
\cite{Ashtekar:1998sp,Ashtekar:1999yj,Ashtekar:2000hw,Ashtekar:2001is}.
The set-up is a variational problem in a portion of spacetime bounded
inside by an axisymmetric weakly isolated horizon $(\Delta,[\ell^a])$
(the extension to multiple horizons is straightforward).

Before talking about the first law, we need to first have suitable
notions of the horizon angular momentum and mass.  We begin with
angular momentum.  Fix an axial symmetry $\varphi^a$ at the
horizon. This means that $\varphi^a$ must preserve (i) the equivalence
class $[\ell^a]$ of null normals that is prescribed for the weakly
isolated horizon, (ii) the intrinsic metric $q_{ab}$ and (iii) the
1-form $\omega_a$.  Furthermore, $\varphi^a$ should commute with
$\ell^a$ and it should look like a rotational vector field in that it
should have closed integral curves, an affine length of $2\pi$, and
should vanish at exactly two null horizon generators.  Consider then a
rotational vector field $\phi^a$ in spacetime such that at $\Delta$ it
is equal to the fixed symmetry: $\left.\phi^a\right|_\Delta =
\varphi^a$.  At infinity, we require that it approach some fixed
rotational symmetry of the asymptotic flat metric.  We then need to
find the Hamiltonian\footnote{Recall that in a phase space, a
  Hamiltonian is responsible for generating time evolution via the
  Poisson bracket.  Thus, for any function $F$ in a phase space,
  $\dot{F} = \{H,F\}$.  In the present case, the phase space consists
  of gravitational (and other) fields which satisfy the appropriate
  boundary conditions. } $H_\phi$ which generates motions along
$\phi^a$.  The Hamiltonian can be shown to reduce to integrals over
the boundaries at the horizon and at infinity.  The term at $\Delta$
is identified with the angular momentum of $\Delta$:
\begin{equation}
  \label{eq:angmomih}
  J_\phi^\Delta = -\frac{1}{8\pi}\oint_S \varphi^a\omega_a\,{}^2\epsilon \,.
\end{equation}
Similarly, the notion of energy corresponds to time translations.
Thus, we consider time evolution vector fields $t^a$ on spacetime such
that at the horizon, it approaches a general symmetry vector $A\ell^a
+ \Omega\varphi^a$.  Here the coefficients $A$ and $\Omega$ are
constants on $\Delta$, but are allowed to vary in phase space.  The
strategy is then again to compute the surface term at $\Delta$ in the
Hamiltonian $H_t$ which generates motions along $t^a$, and the surface
term at $\Delta$ is to be identified as the energy $E_t^\Delta$.
Surprisingly, it turns out that motions along $t^a$ are not always
Hamiltonian and in fact, the Hamiltonian exists if and only if
\begin{equation}
  \label{eq:ihfirstlaw}
  \delta E_t^\Delta = \frac{\kappa_{t}}{8\pi}\delta a_\Delta 
  + \Omega_{t}\delta J_\Delta \,.
\end{equation}
This is just the first law, and a vector field $t^a$ is said to be
\emph{permissible} if the first law for $E_t^\Delta$ is satisfied.  If
the first law is satisfied, it is easy to see that $\kappa_t,\Omega_t$
can depend only on the horizon quantities $(a_\Delta,J_\Delta)$, and
must satisfy integrability condition
\begin{equation}
  \frac{\partial \kappa_t(a_\Delta,J_\Delta)}{\partial J_\Delta} 
  = \frac{\partial \Omega(a_\Delta,J_\Delta)}{\partial a_\Delta}\,.
\end{equation}
This ensures that the right-hand-side of Eq.~(\ref{eq:ihfirstlaw}) can
be integrated to yield an exact variation, and to thus have a well
defined $E_t^\Delta$.

A preferred choice of $t^a$ at the horizon can be obtained by choosing
$\kappa(a_\Delta,J_\Delta)$ and $\Omega(a_\Delta,J_\Delta)$ to have
the same functional dependence on $(a_\Delta,J_\Delta)$ as in the Kerr
spacetime:
\begin{equation}
  \kappa = \frac{R_\Delta^4 - 4J_\Delta^2}{2R_\Delta^3\sqrt{R_\Delta^4 + 4J_\Delta^2}}\,,
  \qquad \Omega = \frac{2J_\Delta}{R_\Delta\sqrt{R_\Delta^4 + 4J_\Delta^2}}\,.
\end{equation}
This leads to the horizon mass:
\begin{equation}
  M_\Delta = \frac{1}{2R_\Delta}\sqrt{R_\Delta^4 + 4J_\Delta^2}\,.
\end{equation}
Finally, we note that $J_{\varphi}^\Delta$ is independent of the choice
of cross-section $S$ and even though it requires a weakly isolated
horizon to carry out the Hamiltonian computation, the formula of
Eq.~(\ref{eq:angmomih}) itself is well defined even on a non-expanding
horizon.  If a cross-section $S$ of $\Delta$ is contained within a
spatial hyper-surface $\Sigma$, and if $\widehat{r}^a$ is the unit spacelike
normal to $S$ in $\Sigma$ and $K_{ab}$ is the extrinsic curvature of
$\Sigma$, then we can rewrite $J_\varphi^\Delta$ as
\begin{equation}
  \label{eq:angmomih2}
  J_\Delta^\varphi = \frac{1}{8\pi}\oint_S K_{ab}\varphi^a \widehat{r}^b\, {}^2\epsilon\,.
\end{equation}
This is particularly convenient in numerical relativity where one
routinely located MTSs on spatial Cauchy surfaces, and would like to
use them to characterize the properties of a black hole in real time
while the simulation is in progress \cite{Dreyer:2002mx}.  The
computation of angular momentum is a surface integral over the MTS and
the mass is just an algebraic expression. These methods are now in
common use in numerical simulations.

\section{Dynamical horizons}
\label{sec:dh}

\subsection{The area increase law}
\label{subsec:dharea} 

The second law for event horizons states that the area of an event
horizon can never decrease in time.  This was first suggested by
Bekenstein and is the starting point for associating the area of a
black hole horizon with entropy.  We have thus far not talked about
the area increase law for quasi-local horizons except in the context
of isolated horizons where it is essentially trivial.  For a dynamical
horizon, the area increase law follows easily from the fact that
$\Theta_{(n)}<0$.  Let $\mathcal{H}$ be a dynamical horizon and $S$ a
generic MTS on it.  Let $r^a$ be the spacelike unit normal to $S$
within $\mathcal{H}$; this is not to be confused with the unit normal
$\widehat{r}^a$ to $S$ which lies on a spatial hyper-surface intersecting
$\mathcal{H}$ as in Fig.~\ref{fig:mtt}.  Let $\tau^a$ be the
unit-timelike vector normal to $\mathcal{H}$.  We assume also that
$r^a$ points outwards, in the sense that if $\widehat{r}^a$ is the
outward normal to $S$ on a spatial hyper-surface $\Sigma$, then
$r^a\widehat{r}_a > 0$.  Finally, let $\mathcal{H}$ be bounded by the
cross-sections $S_1$ and $S_2$.  Then, a suitable choice for the out-
and in-going null normals to $S$ are
\begin{equation}
  \ell^a = \frac{\tau^a + r^a}{\sqrt{2}}\,,\qquad \tau^a = \frac{\tau^a-r^a}{\sqrt{2}}\,.  
\end{equation}
Then, if $q_{ab}$ is the intrinsic Riemannian metric on $S$:
\begin{equation}
  \sqrt{2}q^{ab}\nabla_a r_b = q^{ab}\nabla_a(\ell^a - \tau^a) = \Theta_{(\ell)} - \Theta_{(n)} > 0\,.
\end{equation}
Thus, the area element on $S$ and thus its area increases along $r^a$.
(Though we shall not pursue this further, one could imagine relaxing
the requirement $\Theta_{(n)}<0$ by an average on $S$ and still obtain
the same result).  This is the area increase law for dynamical
horizons.

We can go further and ask whether it is possible to obtain a
\emph{physical process} version of the area increase law which relates
the increase in area from an initial cross-section $S_1$ to a later
time $S_2$ to the amount of energy or radiation falling into the black
hole between $S_1$ and $S_2$.  An early result along these lines was
proved by Hartle and Hawking in 1972 \cite{Hawking:1972hy} for
perturbations of the event horizon.  We note however that such a law
does not exist for event horizons in general. A case in point being
the Vaidya solution which as we have seen, grows in flat space when
nothing falls into the black hole.  Let us then consider the area
increase law on a dynamical horizon $\mathcal{H}$.  Let us first
consider angular momentum and the change in angular momentum from
$S_1$ to $S_2$.  

Since $\mathcal{H}$ is a spacelike hyper-surface we have, as discussed
earlier, the induced metric $h_{ab}$, the associated derivative
operator $D_a$, and the extrinsic curvature $K_{ab}$.  As before, let
$\tau^a$ be the unit timelike normal to $\mathcal{H}$ and let $r^a$ be
the unit spacelike normal to a cross-section $S$ within
$\mathcal{H}$. The Einstein equations then show that $(h_{ab},K_{ab})$
cannot be specified freely, but must satisfy the Hamiltonian and
momentum constraint equations (see e.g. \cite{Wald:1984}).  The
momentum constraint is:
\begin{equation}
  D_b(K^{ab} - Kh^{ab}) = 8\pi T^{bc}n_c h^a_b\,.
\end{equation}
Contracting both sides with a rotational vector field $\varphi^a$ and
integrate by parts to obtain
\begin{equation}
  \label{eq:angmombalance}
  \frac{1}{8\pi}\oint_{S_2} K_{ab}\varphi^ar^b d^2V - \frac{1}{8\pi}\oint_{S_1} K_{ab}\varphi^ar^b = \int_{\mathcal{H}} \left(T_{ab}\tau^a\varphi^b + \frac{1}{16\pi}P^{ab}\Lie_\varphi h_{ab} \right)\,,
\end{equation}
where $P^{ab} = K^{ab} -Kh^{ab}$.  We then identify the angular
momentum of a cross-section $S$ as
\begin{equation}
  \label{eq:angmomdh}
  J_S^{(\varphi)} = -\frac{1}{8\pi}\int_S K_{ab}\varphi^ar^b\,d^2V\,.
\end{equation}
Eq.~(\ref{eq:angmombalance}) is thus a \emph{balance} law for angular
momentum \cite{Ashtekar:2002ag,Ashtekar:2003hk}. It relates
$J_{S_2}^{(\varphi)} - J_{S_1}^{(\varphi)}$ with the gravitational and
matter flux crossing the horizon between $S_2$ and $S_1$.  We note
that if $\varphi^a$ is a Killing vector of the metric $h_{ab}$, then
the gravitational contribution vanishes identically.

A similar (but more involved calculation) leads to a balance law for
the change in area, or rather for the area radius $R$ defined as
$R=\sqrt{a/4\pi}$ \cite{Ashtekar:2002ag,Ashtekar:2003hk}:
\begin{equation}
  \label{eq:areabalance}
  \frac{R_2}{2} - \frac{R_1}{2} = \int_{\mathcal{H}} T_{ab}\tau^a\xi^b d^3V + \frac{1}{16\pi}\int_{\mathcal{H}} N_r\left(|\sigma|^2 + 2|\zeta|^2\right)\,d^3V\,.
\end{equation}
Here $|\sigma|^2:= \sigma_{ab}\sigma^{ab}$ with $\sigma_{ab}$ being
the shear of the outgoing null vector $\ell^a = \tau^a+r^a$;
$|\zeta|^2 := \zeta_a\zeta^b$ where $\zeta^a =
h^{ab}r^c\nabla_c\ell_b$; and $N_r := |D_aR D^a R|^{1/2}$.
Eq.~(\ref{eq:areabalance}) is the desired area balance law.  Again the
flux terms consist of a matter and gravitational contributions.  The
integrand in the gravitational part is manifestly local and
non-negative, and the matter contribution is positive if the dominant
energy condition holds.  The gravitational contribution in fact
vanishes in spherical symmetry as it should.  Along similar lines,
Jaramillo and Gourgoulhon obtained a second order differential
equation for the area \cite{Gourgoulhon:2006uc} which leads to a
causal evolution of the area subject to initial conditions (the
analogous evolution of the area of an event horizon
\cite{Damour:membrane} turns out to be teleological in the sense that
it requires one to specify a boundary condition near future timelike
infinity).

Let us conclude with few words about angular momentum.  We have
identified Eq.~(\ref{eq:angmomdh}) with the angular momentum of a
cross-section of the dynamical horizon, while earlier we had
identified Eq.~(\ref{eq:angmomih2}) with the angular momentum of a
cross-section of an isolated horizon.  It can be shown that when
$\varphi^a$ is a symmetry of the intrinsic metric, then the two agree.
Recently, it was shown by Jaramillo et al \cite{Jaramillo:2011pg},
that for an axisymmetric MOTS which is stably outermost, and if the
spacetime satisfies the dominant energy condition, then the angular
momentum $J$ defined as above, satisfies the inequality $|J| \leq
a/8\pi$ where $a$ is the area of $S$.  Here, axisymmetry is imposed
only at $S$ and not globally.  This surprising result further
validates the identification of Eq.~(\ref{eq:angmomdh}) as the angular
momentum.

\subsection{Uniqueness results for dynamical horizons}
\label{subsec:dhuniqueness}  

In earlier sections, we have already discussed the location and time
evolution of trapped and marginally trapped surfaces.  We now discuss
further results obtained by Ashtekar and Galloway
\cite{Ashtekar:2005ez} related to the issue of uniqueness of
marginally trapped surfaces in the dynamical case.

The first result, proved in \cite{Ashtekar:2005ez}, concerns the
uniqueness of the foliation of a dynamical horizon by marginally
trapped surfaces.  Let $\mathcal{H}$ be a dynamical horizon foliated
by a set marginally trapped surfaces $S_t$ with $t$ being a continuous
real parameter taking values within an open interval.  Let $S$ be a
\emph{weakly} trapped surface in $\mathcal{H}$, i.e. both of its
expansions are non-positive: $\Theta_{(\ell)}\leq 0$,
$\Theta_{(n)}\leq 0$ (in particular, $S$ could be a marginally trapped
surface).  Then $S$ must in fact be a marginally trapped surface and
must coincide with one of the $S_t$.  This shows that the foliation of
$\mathcal{H}$ by the $S_t$ must be unique.  As a corollary, consider
as in Fig.~\ref{fig:mtt}, a MTT generated by MTSs $S_t$ which lie in
spacelike surfaces $\Sigma_t$.  This is a situation common in
numerical relativity where one uses the $\Sigma_t$ for time evolution,
and one locates MTSs on them.  If we were to use a different
foliation, and in particular a particular spatial hyper-surface
$\Sigma^\prime$ which does not coincide with any of the $\Sigma_t$.
If the intersection $\Sigma^\prime\bigcap\mathcal{H}$ is not one of
the $S_t$, then it cannot be a marginally trapped surface.  This is
illustrated in Fig.~\ref{fig:vaidya-numrel} for the Vaidya
spacetime. The intersection of the non-symmetric spatial hyper-surface with
the spherically symmetric dynamical horizon is the red surface, and it
is not a marginally (or weakly) trapped surface.  Thus, if we had a
different foliation $\Sigma^\prime_{t^\prime}$ and we locate MTSs on
these spatial hyper-surfaces, then we would end up with a \emph{different}
dynamical horizon $\mathcal{H}^\prime$.  We note again the dramatic
difference for an isolated horizon where every spherical cross-section
is a marginally outer trapped surface.

The question then arises, how different can $\mathcal{H}^\prime$ be
from $\mathcal{H}$.  A partial answer is provided by the next result
\cite{Ashtekar:2005ez}: \emph{There are no weakly trapped surfaces
  contained in the past domain of dependence $D^-(\mathcal{H})$ (apart
  from those which make up $\mathcal{H}$ itself)}\footnote{The past
  domain of dependence $D^-(\mathcal{H})$ is the set of spacetime
  points $p$ such that every future causal curve from $p$ intersects
  $\mathcal{H}$}.  In particular, this result rules out a dynamical
horizon contained in $D^-(\mathcal{H})-\mathcal{H}$. Once again, this
result is illustrated by the Vaidya example discussed earlier in
Fig.~\ref{fig:vaidya}.  The non-symmetric surface $S$ in this figure
is partly inside and partly outside the spherically symmetric
dynamical horizon.  More generally, weakly trapped surfaces lying
partly inside $D^-(\mathcal{H})$ are not ruled out.
\begin{figure}
  \centering
  \includegraphics[width=0.4\textwidth]{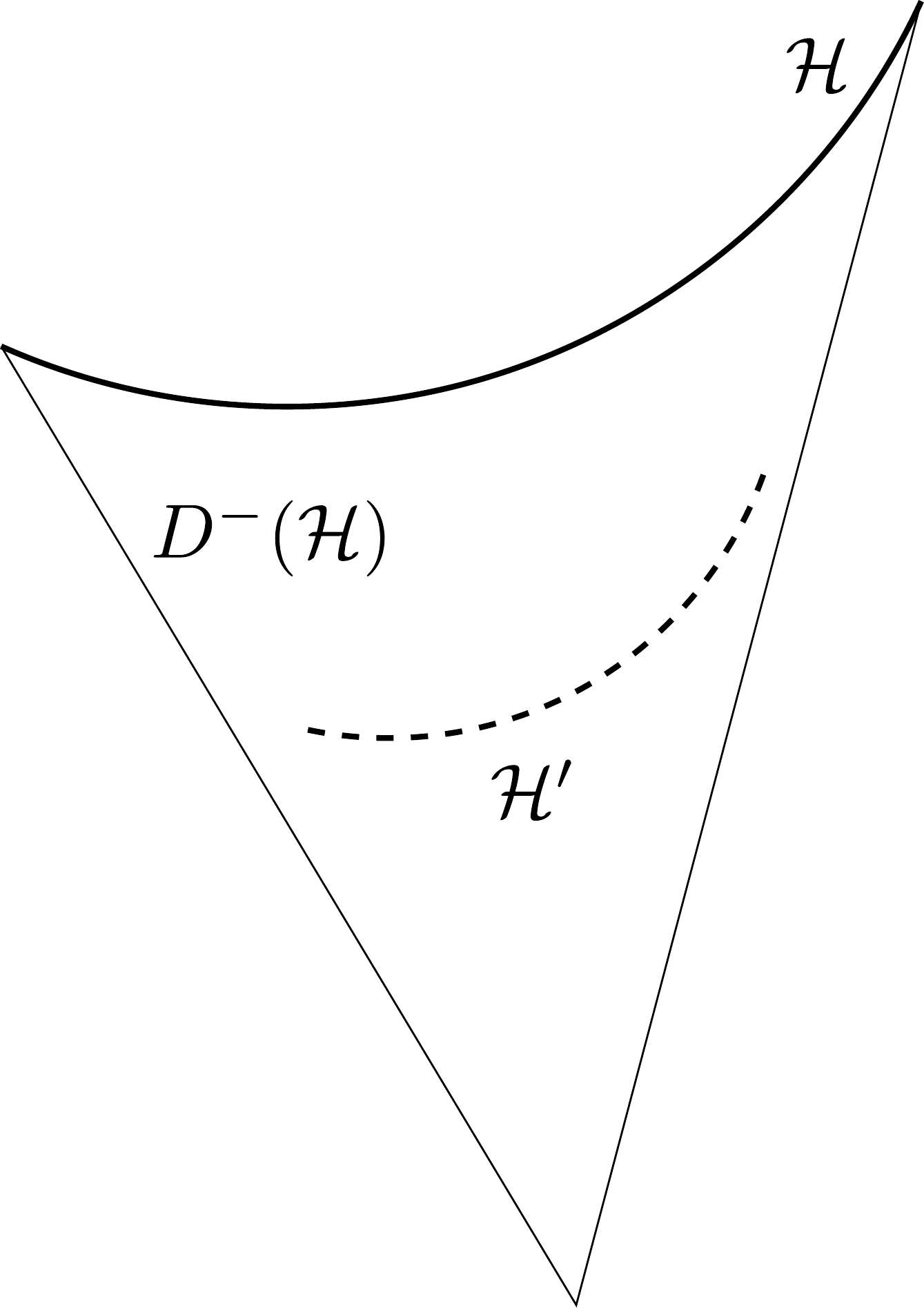}
  \caption{This situation is ruled out by the uniqueness results of
    Ashtekar and Galloway.  If there is a dynamical horizon
    $\mathcal{H}$, then there cannot be another dynamical horizon
    $\mathcal{H}^\prime$ lying completely in the past domain of dependence
    $D^-(\mathcal{H})$. }
  \label{fig:dhuniqueness}
\end{figure}

\section{Outlook}
\label{sec:outlook}

As discussed in the introduction, ever since the discover of the
Schwarzschild solution almost a century ago, research in black holes
has led to seminal developments in theoretical, observational and
computational physics.  This includes the singularity theorems, black
hole thermodynamics, the uniqueness theorems, the cosmic censorship
hypothesis, black hole entropy calculations in various approaches to
quantum gravity, the various astrophysical phenomena involving black
holes, and the recent results from numerical relativity.  The
framework of quasi-local horizons, which takes trapped and marginally
trapped surfaces as its starting point, provides a unified approach
for studying various aspects of black hole physics.  While we have
only touched upon a few aspects and applications of this framework,
the material presented in this chapter will hopefully motivate the
reader to delve further into the subject.  An important theme in this
discussion is that to base our understanding of black holes on lessons
from stationary cases can be misleading.  Dynamical situations have
some essentially different features and intuition from stationary
examples can easily lead us astray.  We have illustrated this by the
Schwarzschild and Vaidya examples.

In this chapter we have introduced trapped surfaces and various kinds
of quasi-local horizons through examples.  The eventual goal of these
studies (from a physics viewpoint) is to understand the properties of
the surface of a black hole.  We have discussed the inadequacy of
event horizons for this purpose due to its teleological properties.
and it is desirable to find a suitable replacement.  Penrose's trapped
surfaces and the boundary of the trapped region seem ideally suited
for this task and lead naturally to the various definitions of
quasi-local horizons.  The simplest example is of course the
Schwarzschild black hole.  In this example, the boundary of the
trapped region agrees with the event horizon, and both notions give
rise to the same physical ideas. Difficulties arise however when we
consider non-stationary black holes.  We illustrated this through the
imploding Vaidya spacetime.  The intuitively obvious horizon, the
analog of the $r=2M$ hyper-surface in Schwarzschild is a spherically
symmetry dynamical horizon and it is separated from the event horizon.
However, the non-spherically symmetric outer trapped surfaces extend
up to the event horizon. To make matters more complicated, trapped
surfaces do not extend all the way to the event horizon.  We used
these examples as motivations for general definitions and we reviewed
some basic results regarding trapped surfaces and quasi-local
horizons.  We saw that marginally trapped surfaces are not as
ill-behaved as one might think, and under physically reasonable
conditions, they do evolve smoothly.  The equilibrium case, described
by isolated horizons is also of great interest.  It covers a wide
variety of situations where a black hole is in equilibrium in a
dynamical spacetime and is the best understood quasi-local horizon.
Finally, in the dynamical case, we saw that one can assign physical
quantities such as mass, angular momentum and fluxes through dynamical
horizons.  

There are a number of topics that we have not discussed.  In
particular, we have not discussed the various applications of these
notions in numerical relativity.  Similarly, our discussion has been
restricted to the classical world and we have not talked about
e.g. the quantization of isolated horizons and the black hole entropy
calculations.  Even in the discussion of the mathematical properties
of quasi-local horizons, we have discussed black hole multipole
moments, symmetries, and inclusion of various kinds of matter fields
only very briefly.  Another significant omission is the discussion of
black holes near equilibrium.  Reviews of these topics can be found in
e.g. \cite{Gourgoulhon:2005ng,Booth:2005qc,Ashtekar:2004cn}

We have seen that the outstanding question in this field is the
non-uniqueness of dynamical horizons. We have discussed some
restrictions on where dynamical horizons can be located.  However, it
is a fact that there are a multitude of smooth marginally trapped
tubes and dynamical horizons in a general black hole spacetime, and
there seems to be no obvious way of picking a preferred one.  While
there could be reasonable choices in specific examples, there does not
seem to be any general solution to this problem. While one can get
away with using event horizons in globally stationary spacetimes,
choosing to live with event horizons in general is not an option
because of its global properties.  One could perhaps consider dealing
with the full set of dynamical horizons and marginally trapped tubes.
Many of them are smooth one can study them individually; e.g. the
dynamical horizon flux formulae apply to all dynamical horizons
equally. However, if we wish to assign physical properties to the
black hole such as mass, angular momentum, fluxes and higher
multipoles, which one should we choose?  We will generally get
different results depending on our choice. The boundary of the trapped
region could be a reasonable alternative, but as we have seen, it is
generally not a marginally trapped tube itself.  It is also not clear
whether one can use this boundary to study, say, black hole
thermodynamics and other physical phenomena that we believe are true
for black holes and besides, this boundary also has a number of global
properties and is difficult to locate (thus making it not better than
event horizons in many regards).  An interesting possibility,
suggested by Bengtsson and Senovilla, is the \emph{core} of the black
hole region, i.e. the portion of the trapped region where all trapped
surfaces penetrate.  Removing the core would then completely eliminate
all trapped surfaces.  If it is indeed a proper subset of the trapped
region, then is its boundary a dynamical horizon?  There are
indications that the region $r\leq 2M(v)$ region of Vaidya is such a
core, and its boundary is the spherically symmetric dynamical
horizon. However the core is not unique and some cores are not
spherically symmetric \cite{Bengtsson:2010tj}.  It is an interesting
open question whether this idea can be developed further.

\section*{Acknowledgment}

I am grateful to Abhay Ashtekar for valuable discussions and
suggestions.

\bibliographystyle{abbrv}
\bibliography{refs}

\end{document}